\documentclass[graybox,vecphys]{svmult}

\usepackage{helvet}         
\usepackage{courier}        
\usepackage{makeidx}         
\usepackage{graphicx}        
\usepackage{multicol}        
\usepackage[bottom]{footmisc}
\ifcsname SQRT \endcsname
\else
 
 \newcommand\LEFTRIGHT[3]{{\left#1 #3 \right#2}}
\fi
\usepackage{cite}

\makeindex             

\begin{document}

\title*{Quantum phase transitions of antiferromagnets\\ and the cuprate superconductors}
\author{Subir Sachdev}
\institute{Subir Sachdev \at Department of Physics, Harvard University, Cambridge MA 02138, USA.\\ \email{sachdev@physics.harvard.edu}}
%
%
\maketitle

\abstract{
I begin with a proposed global phase diagram of the cuprate superconductors as a function
of carrier concentration, magnetic field, and temperature, and highlight its connection to numerous
recent experiments. The phase diagram is then used as a point of departure for a pedagogical
review of various quantum phases and phase transitions of insulators, superconductors,
and metals. The bond operator method is used
to describe the transition of dimerized antiferromagnetic insulators between magnetically ordered states and
spin-gap states. The Schwinger boson method is applied to frustrated square lattice antiferromagnets:
phase diagrams containing collinear and spirally ordered magnetic states, $Z_2$ spin liquids, and valence
bond solids are presented, and described by an effective gauge theory of spinons. Insights from these
theories of insulators are then applied to a variety of symmetry breaking transitions in $d$-wave superconductors. The latter systems also contain fermionic quasiparticles with a massless Dirac spectrum,
and their influence on the order parameter fluctuations and quantum criticality is carefully discussed.
I conclude with an introduction to strong coupling problems associated with symmetry breaking
transitions in two-dimensional metals, where the order parameter fluctuations couple to a gapless line
of fermionic excitations along the Fermi surface.}

\section{Introduction}
\label{sec:intro}

The cuprate superconductors have stimulated a great deal of innovative theoretical
work on correlated electron systems. On the experimental side, new experimental techniques
continue to be discovered and refined, leading to striking advances over twenty years
after the original discovery of high temperature superconductivity \cite{nobel}.

\begin{figure}[t]
\centering
\includegraphics[width=4in]{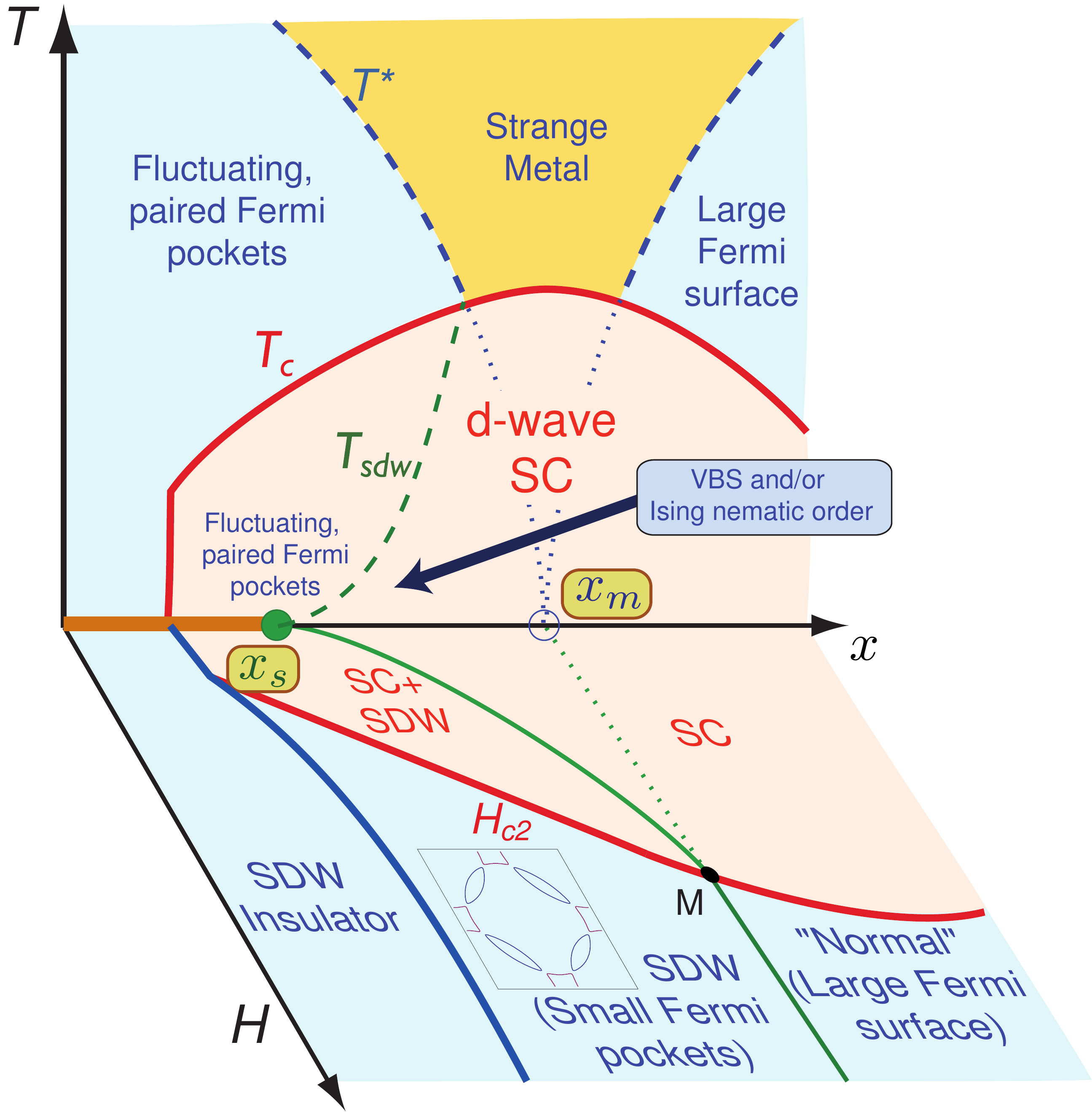}
\caption{Proposed global phase diagram for the hole- and electron-doped cuprates \cite{moon,dresden,m2s}.
The axes are carrier concentration ($x$), temperature ($T$), and a magnetic field ($H$) applied
perpendicular to the square lattice. The regions with the SC label have $d$-wave superconductivity. The strange
metal and the ``pseudogap'' regime are separated by the temperature $T^\ast$. Dashed lines indicate
crossovers. After accounting for the valence bond solid (VBS) or Ising nematic orders
that can appear in the regime $x_s < x < x_m$, the dashed $T^\ast$ line and the dotted line connecting
$x_m$ to the point $M$ become true phase transitions. There can also be fractionalized phases in the region $x_s < x < x_m$ , as discussed recently in Refs.~\cite{newref1,newref2}.} \label{fig:figglobal}
\end{figure}

In the past few years, a number of experiments, and most especially the discovery
of quantum oscillations in the underdoped regime \cite{louis1}, have shed remarkable new light
on the origins of cuprate superconductivity. I believe these new experiments point
to a synthesis of various theoretical ideas, and that a global theory of the rich
cuprate phenomenology may finally be emerging. The ingredients for this synthesis were described in
Refs.~\cite{moon,dresden,m2s}, and
are encapsulated in the phase diagram shown
in Fig.~\ref{fig:figglobal}.
Here I will only highlight a few important features of this phase diagram, and use those
as motivations for the theoretical models described in these lectures. The reader is referred to the earlier
papers \cite{dresden,m2s} for a full discussion of the experimental support for these ideas.
Throughout the lectures, I will refer back to Fig.~\ref{fig:figglobal} and point out the relevance of
various field theories to different aspects of this rich phase diagram.

It is simplest to examine the structure of Fig.~\ref{fig:figglobal} beginning from the regime of large doping.
There, ample evidence has established that the ground state is a conventional Fermi liquid, with a single ``large''
Fermi surface enclosing the area demanded by the Luttinger theorem. Because of the underlying band structure,
this large Fermi surface is hole-like (for both electron and hole-doping), and so encloses an area $1+x$
for hole density $x$, and an area $1-p$ for doped electron density $p$. The central quantum phase
transition (QPT)  in Fig.~\ref{fig:figglobal} is the onset of spin density wave (SDW)
order in this large Fermi surface metal
at carrier concentration $x=x_m$, shown in Fig.~\ref{fig:figglobal} near the region 
where $T_c$ is largest (the subscript $m$ refers to the fact that the transition takes place in a metal);
we will describe this transition in more detail in Section~\ref{sec:metal}.
Because of the onset of superconductivity, the QPT at $x=x_m$ is revealed only at magnetic fields
strong enough to suppress superconductivity, {\em i.e.\/}, at $H>H_{c2}$. For $x<x_m$, we then have
a Fermi liquid metal with SDW order. Close to the transition, when the SDW order is weak, the large Fermi surface
is generically broken up by the SDW order into ``small'' electron and hole pockets, each enclosing an area
of order $x$ (see Fig.~\ref{fig:sdw} later in the text).
Note that electron pockets are present {\em both\/} for hole and electron doping: such electron
pockets in the hole-doped cuprates were first discussed in Ref.~\cite{scs}. There is now
convincing experimental evidence for the small Fermi pockets in the hole underdoped cuprates,
including accumulating evidence for electron pockets \cite{louis2,amro}. The QPT between the small
and large Fermi surface metals is believed to be at $x_m \approx 0.24$ in the hole-doped
cuprates \cite{nlsco1,louisnematic}, and at $p_m \approx 0.165$ in the electron-doped cuprates \cite{helm1,helm2}.
One of the central claims of  Fig.~\ref{fig:figglobal} is that it is the QPT at $x=x_m$ which controls the non-Fermi liquid
``strange metal'' behavior in the normal state above the superconductivity $T_c$. We leave open the
possibility \cite{su2acl} that there is an extended non-Fermi liquid phase
for a range of densities with $x>x_m$: this is not shown in Fig.~\ref{fig:figglobal}, and
will be discussed here only in passing.

The onset of superconductivity near the SDW ordering transition of a metal has been considered in numerous
previous works \cite{scalapino,acs}. These early works begin with the large Fermi surface found for $x>x_m$,
and consider pairing induced by exchange of SDW fluctuations; for the cuprate Fermi surface geometry,
they find an attractive interaction in the $d$-wave channel, leading to $d$-wave superconductivity. Because
the pairing strength is proportional to the SDW fluctuations, and the latter increase as $x \searrow x_m$,
we expect $T_c$ to increase as $x$ is decreased for $x>x_m$, as is shown in Fig.~\ref{fig:figglobal}.
Thus for $x>x_m$, stronger SDW fluctuations imply stronger superconductivity, and the orders effectively
attract each other.

It was argued in Ref.~\cite{moon} that the situation becomes qualitatively different for $x<x_m$.
This becomes clear from an examination of Fig.~\ref{fig:figglobal} as a function of decreasing $T$ for
$x<x_m$. It is proposed \cite{galitski,moon} in the figure that the Fermi surface already breaks
apart locally into the small pocket Fermi surfaces for $T<T^\ast$. So the onset of superconductivity
at $T_c$ involves the pairing of these small Fermi surfaces, unlike the large Fermi surface pairing
considered above for $x>x_m$. For $x<x_m$, an increase in local SDW ordering is not conducive
to stronger superconductivity: the SDW order `eats up' the Fermi surface, leaving less room
for the Cooper pairing instability on the Fermi surface. Thus in this regime we find a {\em competition\/}
between SDW ordering and superconductivity for `real estate' on the Fermi surface \cite{machida,moon}. As we expect the SDW
ordering to increase as $x$ is decreased for $x<x_m$, we should have a decrease in $T_c$
with decreasing $x$, as is indicated in Fig.~\ref{fig:figglobal}.

We are now ready to describe the second important feature of Fig.~\ref{fig:figglobal}.
The complement of the suppression of superconductivity by SDW ordering is the suppression of SDW
ordering by superconductivity.
The competition between superconductivity and SDW order moves \cite{moon}
the actual SDW onset at $H=0$ and $T=0$ to a lower carrier concentration
$x=x_s$ (or $p=p_s$ for electron doping). The QPT at $x=x_s$ controls the criticality of spin fluctuations
within the superconducting phase (and hence the subscript $s$), while that $x=x_m$ continues to be important
for $T>T_c$ (as is indicated in Fig.~\ref{fig:figglobal}).  There is now a line of SDW-onset transitions
within the superconducting phase \cite{demler},
connecting the point $x_s$ to the point $M$, for which there is substantial
experimental evidence \cite{lake1,lake2,boris,mesot1,mesot2,keimer}. The magnitude of the shift from $x_m$ to $x_s$ depends a great
deal upon the particular cuprate: it is largest in the materials with the strongest superconductivity and the highest $T_c$.
In the hole-doped YBCO series we estimate $x_s \approx 0.085$ \cite{keimer} and
in the hole-doped LSCO series we have $x_s \approx 0.14$ \cite{boris} (recall our earlier
estimate $x_m \approx 0.24$ in the hole-doped
cuprates \cite{nlsco1,louisnematic}),
while in the electron-doped cuprate
Nd$_{2-x}$Ce$_x$CuO$_4$, we have $p_s = 0.145$ \cite{greven}
(recall $p_m \approx 0.165$ in the electron-doped cuprates \cite{helm1,helm2}).

With the shift in SDW ordering from $x_m$ to $x_s$, the need for the crossover line labeled
$T_{\rm sdw}$ in Fig.~\ref{fig:figglobal} becomes evident. This is the temperature at which the
electrons finally realize that they are to the `disordered' of the actual SDW ordering transition
at $x=x_s$, rather than to the `ordered' side of the transition at $x=x_m$. Thus, for $T<T_{\rm sdw}$,
the large Fermi surface re-emerges at the lowest energy scales, and SDW
order is never established. This leaves us with an interesting superconducting state at $T=0$, where
the proximity to the Mott insulator can play an important role. Other orders linked to the antiferromagnetism
of the Mott insulator can appear here, such as valence bond solid (VBS) and Ising-nematic order \cite{edoped}, or even topologically 
ordered phases \cite{newref1,newref2};
experimental evidence for such orders has appeared in a number of recent
experiments \cite{kohsaka1,ando02,hinkov08a,louisnematic}, and
we will study these orders in the sections below.

\begin{figure}[t]
\centering
\includegraphics[width=2.25in]{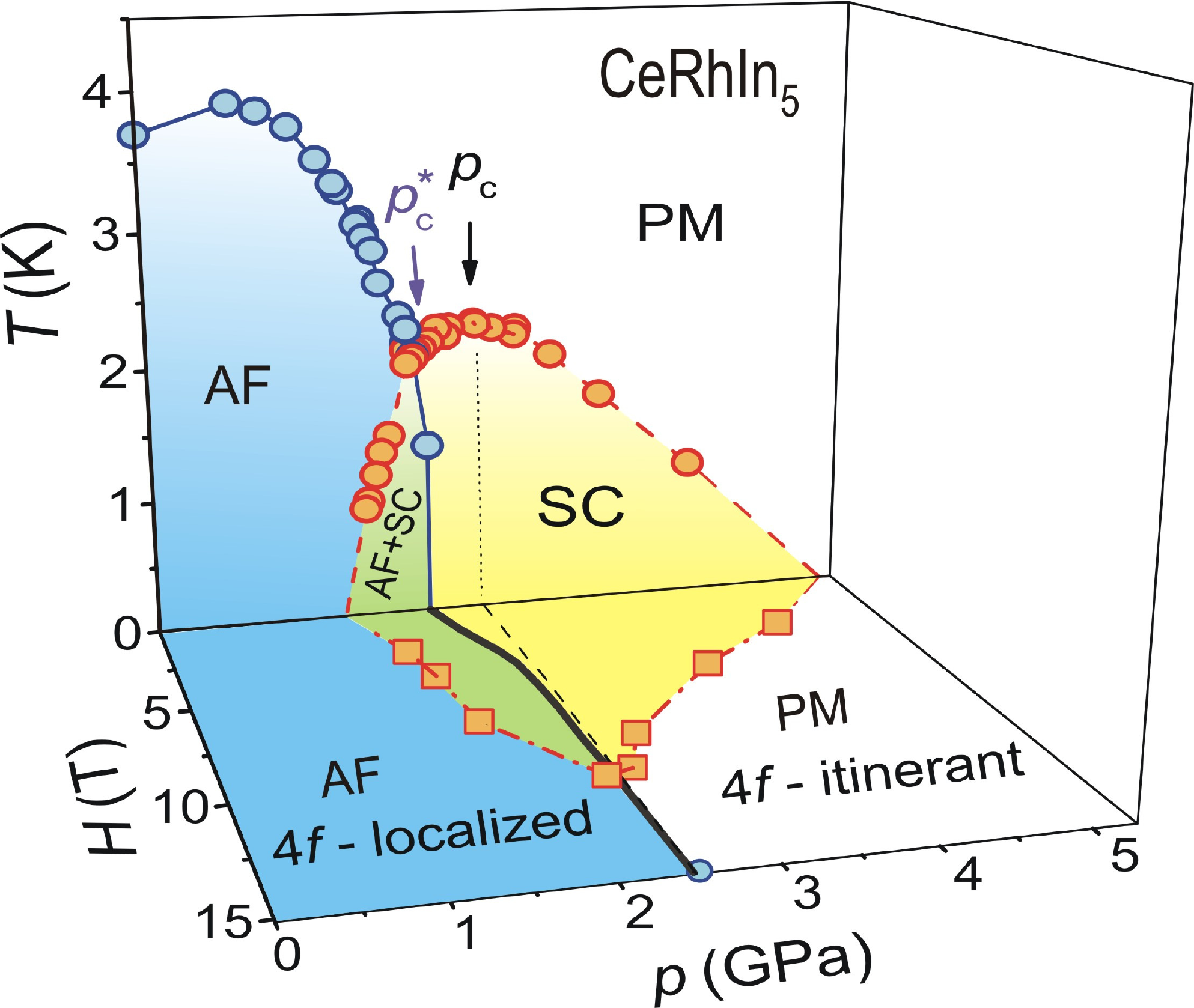}
\includegraphics[width=2.25in]{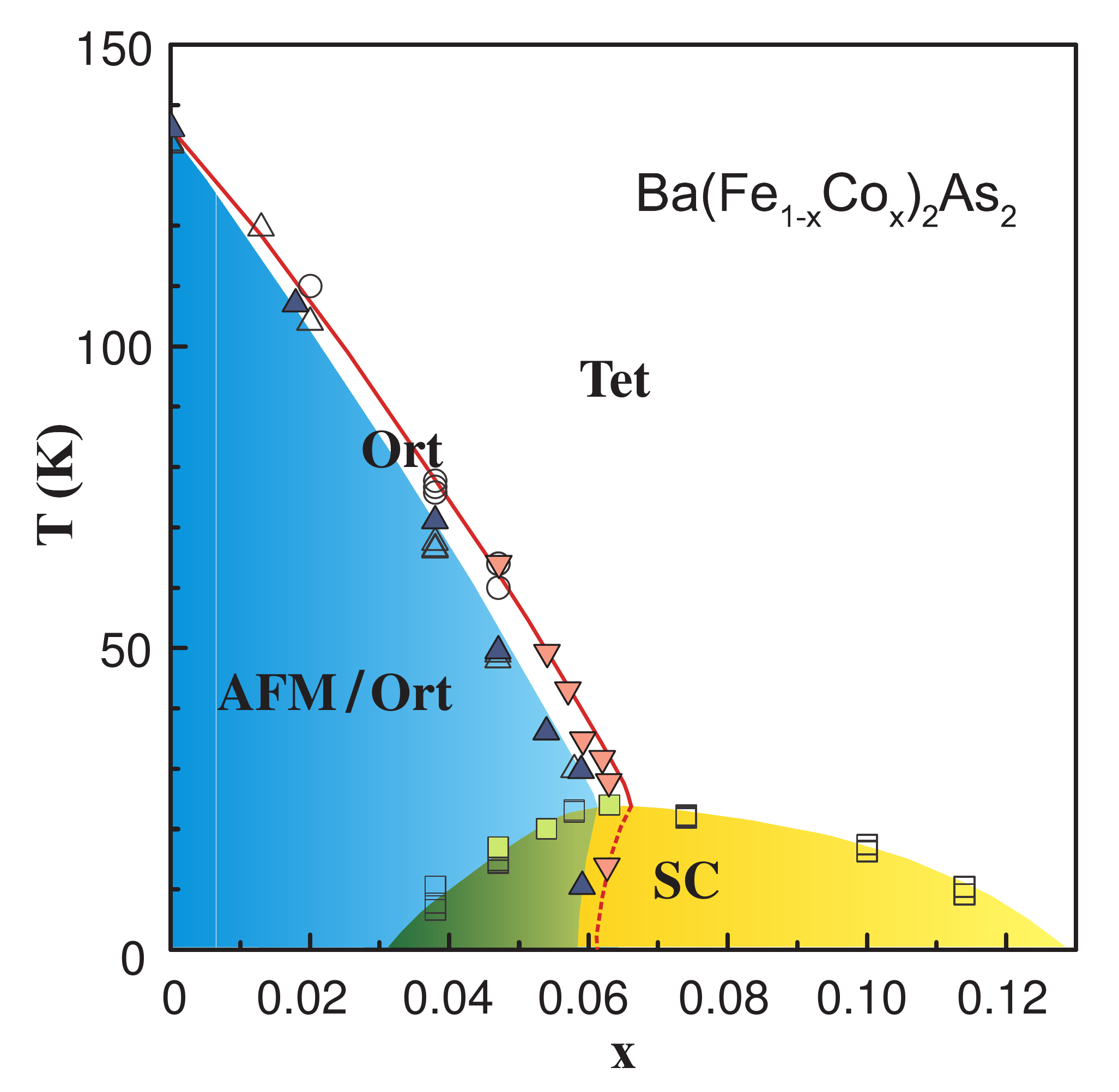}
\caption{Phase diagrams for CeRhIn$_5$ from Ref.~\cite{knebel} and for Ba[Fe$_{1-x}$Co$_x$]$_2$As$_2$
from Refs.~\cite{joerg0,joerg1,joerg2}. For CeRhIn$_5$, the shift from $p_c$ to $p_c^\ast$ is similar
to the shift from $x_m$ to $x_s$ in Fig.~\ref{fig:figglobal}; this shift is significantly larger in the cuprates (and
especially in YBCO) because
the superconductivity is stronger. In the ferropnictide Ba[Fe$_{1-x}$Co$_x$]$_2$As$_2$, the back-bending of the
SDW ordering transition in the superconducting phase is similar to that of $T_{\rm sdw}$ in Fig.~\ref{fig:figglobal}.
} \label{fig:knebel}
\end{figure}

The shift in the SDW ordering from $x_m$ to $x_s$ has recently emerged as a generic property
of quasi-two-dimensional correlated electron superconductors, and is not special to the cuprates.
Knebel {\em et al.} \cite{knebel} have presented a phase diagram for CeRhIn$_5$ as a function of temperature, field,
and pressure (which replaces carrier concentration) which is shown in Fig.~\ref{fig:knebel}.
Notice the very similar structure to Fig.~\ref{fig:figglobal}:
the critical pressure for the onset of antiferromagnetism shifts from the metal to the superconductor, so that
the range of antiferromagnetism is smaller in the superconducting state. In the pnictides, the striking observations
by the Ames group \cite{joerg0,joerg1,joerg2} on Ba[Fe$_{1-x}$Co$_x$]$_2$As$_2$
show a `back-bending' in the SDW onset temperature
upon entering the superconducting phase: see Fig.~\ref{fig:knebel}. This is similar to the back-bending of the line $T_{\rm sdw}$ from $T^\ast$
in Fig.~\ref{fig:figglobal}, and so can also be linked to the shift in the SDW onset transition between the
metal and the superconductor.

It is clear from Fig.~\ref{fig:knebel} that the shift in the SDW order
between the metal and the superconductor is relatively small in the
non-cuprate materials, and may be overlooked in an initial
study without serious consequences.
Similar comments apply to the electron-doped cuprates. However, the shift is quite large
in the hole-doped cuprates: this can initially suggest that the cuprates are a different class
of materials, with SDW ordering playing a minor role in the physics of the superconductivity.
One of the main claims of Fig.~\ref{fig:figglobal} is that after accounting
for the larger shift in the SDW transition, all of the cuprates fall into a much wider class of correlated
electron superconductors for which the SDW ordering transition in the metal
is the central QPT controlling the entire phase diagram
(see also the recent discussion by Scalapino \cite{scalapinom2s}).

Our discussion will be divided into 3 sections, dealing with the nature of quantum fluctuations
near SDW ordering in insulators, $d$-wave superconductors, and metals respectively.
These cases are classified according to the increasing density of states for single-electron excitations.
We will begin in Section~\ref{sec:insulator} by considering a variety of Mott insulators, and describe
their phase diagrams. The results will apply directly to experiments on insulators not part of the
cuprate family. However, we will also gain insights, which will eventually be applied to various
aspects of Fig.~\ref{fig:figglobal} for the cuprates. Then we will turn in Section~\ref{sec:dwave} to $d$-wave
superconductors, which have a Dirac spectrum of single-electron excitations as described in Section~\ref{sec:dirac}.
Their influence
on the SDW ordering transition at $x=x_s$ will be described using field-theoretical methods in Section~\ref{magd}.
Section~\ref{sec:ising} will describe the Ising-nematic ordering at or near $x=x_s$ indicated in Fig.~\ref{fig:figglobal}.
Finally, in Section~\ref{sec:metal}, we will turn to metals, which have a Fermi
surface of low-energy
single-particle excitations. We will summarize the current status of QPTs of metals: in two dimensions
most QPTs lead to strong coupling problems which have not been conquered. It is clear from Fig.~\ref{fig:figglobal}
that such QPTs are of vital importance to the physics of metallic states near $x=x_m$.

Significant portions of the discussion in the sections below have been adapted from
other review articles by the author \cite{trieste,qpmott,altenberg}.

\section{Insulators}
\label{sec:insulator}

The insulating state of the cuprates at $x=0$ is a $S=1/2$ square lattice antiferromagnet,
which is known to have long-range N\'eel order. We now wish to study various routes by
which quantum fluctuations may destroy the N\'eel order. In this section, we will do this
by working with undoped insulators in which we modify the exchange interactions. These do not
precisely map to any of the transitions in the phase diagram in Fig.~\ref{fig:figglobal}, but we will see in
the subsequent sections that closely related theories do play an important role.

The following subsections will discuss two distinct routes to the destruction of N\'eel order
in two-dimensional antiferromagnet. In Section~\ref{sec:dimer} we describe coupled dimer
antiferromagnets, in which the lattice has a natural dimerized structure, with 2 $S=1/2$
spins per unit cell which can pair with each other. These are directly relevant to experiments on materials
like TlCuCl$_3$. We will show that these antiferromagnets can be efficiently described
by a bond-operator method.
Then in section~\ref{sec:frust} we will consider the far more complicated and subtle case
where the lattice has full square lattice symmetry with only a single $S=1/2$ spin per unit cell, and the N\'eel
order is disrupted by frustrating exchange interactions. We will explore the phase diagram of such antiferromagnets
using the Schwinger boson method. These results have direct application to
experimental and numerical studies of a variety
of two-dimensional Mott insulators on the square, triangular, and kagome lattices; such
applications have been comprehensively reviewed in another recent
article by the author \cite{solvay}, and so will not be repeated here.

\subsection{Coupled dimer antiferromagnets: bond operators}
\label{sec:dimer}

\begin{figure}[t]
\centering
\includegraphics[width=2.5in]{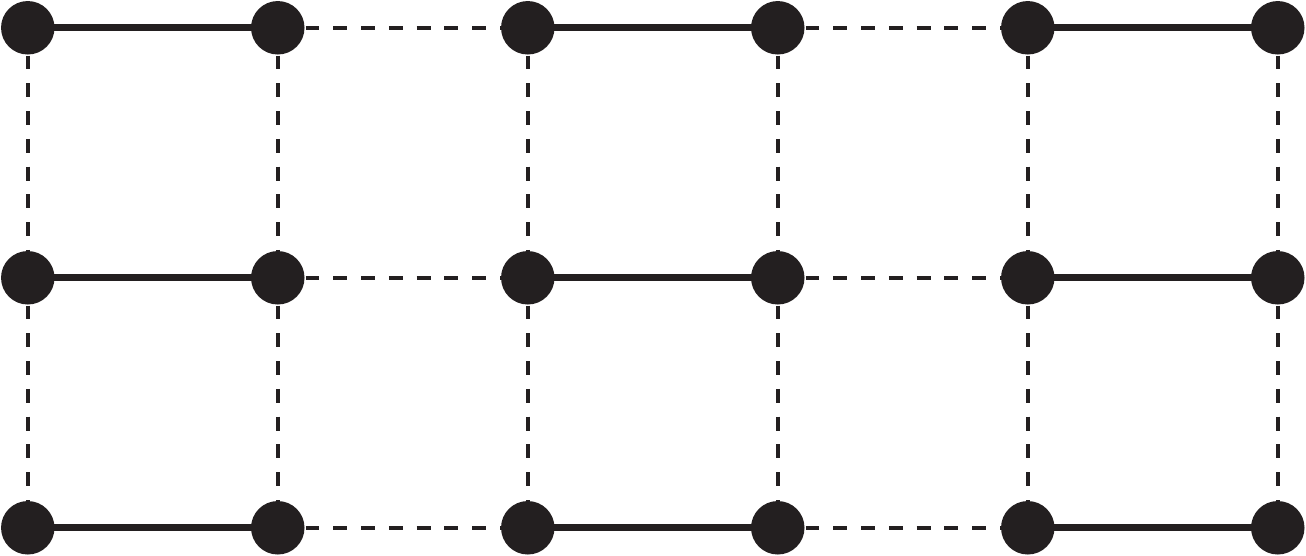}
\caption{ The coupled dimer antiferromagnet. Spins ($S=1/2$) are
placed on the sites, the $\mathcal{A}$ links are shown as full
lines, and the $\mathcal{B}$ links as dashed lines.} \label{fig1}
\end{figure}

We consider the ``coupled dimer'' Hamiltonian \cite{gsh}
\begin{equation}
H_{d} = J \sum_{\langle ij\rangle \in \mathcal{A}} \vec{S}_i \cdot
\vec{S}_j + g J \sum_{\langle ij \rangle \in \mathcal{B}}
\vec{S}_i \cdot \vec{S}_j \, , \label{ham}
\end{equation}
where $\vec{S}_j$ are spin-1/2 operators on the sites of the
coupled-ladder lattice shown in Fig.~\ref{fig1}, with the
$\mathcal{A}$ links forming decoupled dimers while the
$\mathcal{B}$ links couple the dimers as shown.
The ground state of $H_d$ depends only on the dimensionless
coupling $g$, and we will describe the low temperature ($T$)
properties as a function of $g$. We will restrict our
attention to $J>0$ and $0 \leq g \leq 1$. A three-dimensional
model with the same structure as $H_d$ describes the insulator
TlCuCl$_3$ \cite{oosawa,ruegg,ssruegg}.

Note that exactly at $g =1$, $H_d$ is identical to the
square lattice antiferromagnet, and this is the only point at
which the Hamiltonian has only one spin per unit cell. At all
other values of $g$, $H_d$ has a pair of $S=1/2$ spins in
each unit cell of the lattice.

\subsubsection{Phases and their excitations}
\label{sec:lad1}

Let us first consider the case where $g$ is close to 1.
Exactly at $g=1$, $H_d$ is identical to the square lattice
Heisenberg antiferromagnet, and this is known to have long-range,
magnetic N\'{e}el order in its ground state, {\em i.e.}, the
spin-rotation symmetry is broken and the spins have a non-zero,
staggered, expectation value in the ground state with
\begin{equation}
\langle \vec{S}_j \rangle = \eta_j N_0 \vec{n}, \label{neel}
\end{equation}
where $\vec{n}$ is some fixed unit vector in spin space, $\eta_j$
is $\pm 1$ on the two sublattices, and $N_0$ is the N\'{e}el order
parameter. This long-range order is expected to be preserved for a
finite range of $g$ close to 1. The low-lying excitations
above the ground state consist of slow spatial deformations in the
orientation $\vec{n}$: these are the familiar spin waves, and they
can carry arbitrarily low energy, {\em i.e.}, the phase is
`gapless'. The spectrum of the spin waves can be obtained from a
text-book analysis of small fluctuations about the ordered
N\'{e}el state using the Holstein-Primakoff method
\cite{callaway}: such an analysis yields {\em two} polarizations
of spin waves at each wavevector $k = (k_x, k_y)$ (measured from
the antiferromagnetic ordering wavevector), and they have
excitation energy $\varepsilon_k = (c_x^2 k_x^2 + c_y^2
k_y^2)^{1/2}$, with $c_x, c_y$ the spin-wave velocities in the two
spatial directions.

\begin{figure}[t]
\centering
\includegraphics[width=2.5in]{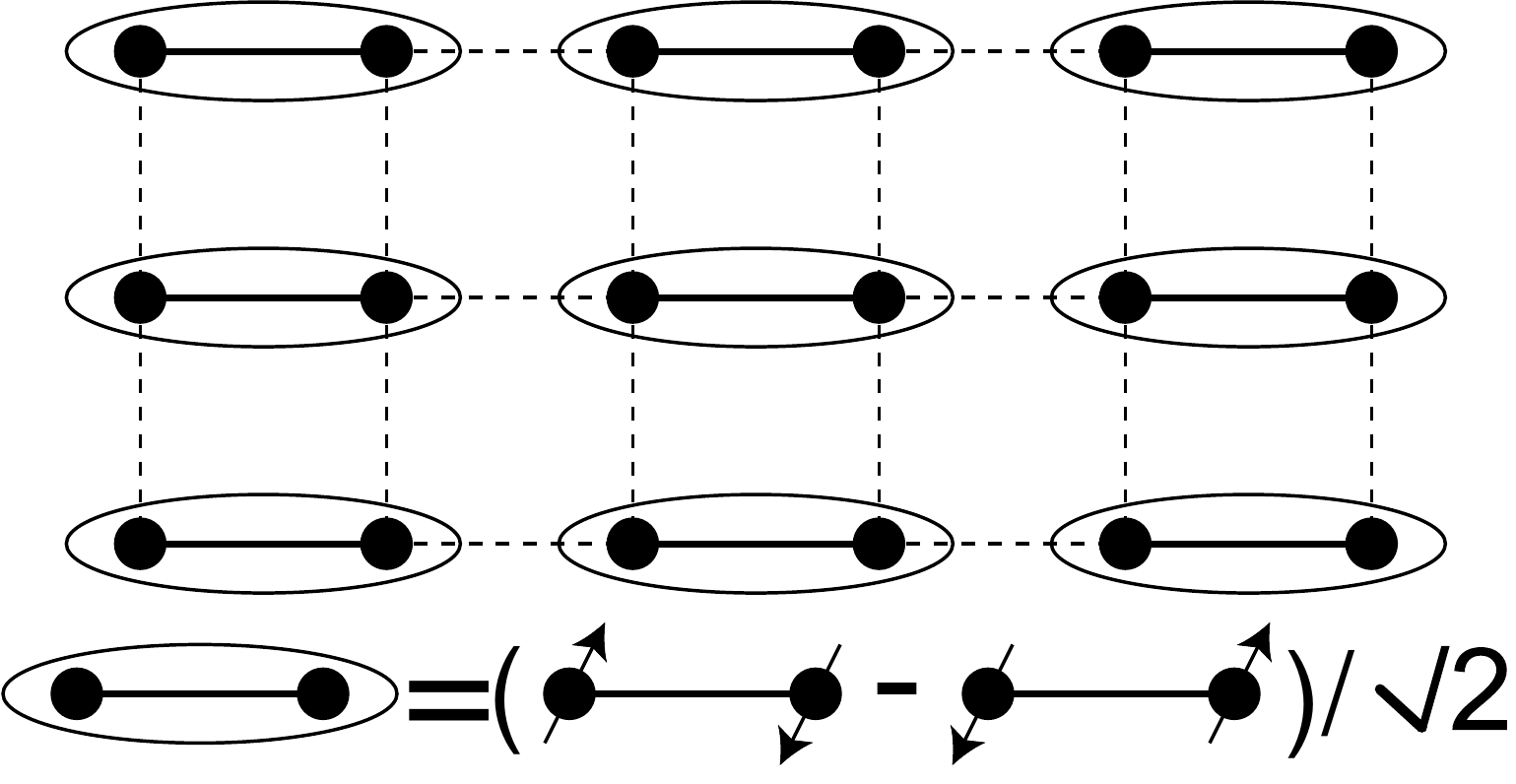}
\caption{Schematic of the quantum paramagnet ground state for
small $g$. The ovals represent singlet valence bond pairs.
}\label{fig2}
\end{figure}

\begin{figure}[t]
\centering
\includegraphics[width=4.5in]{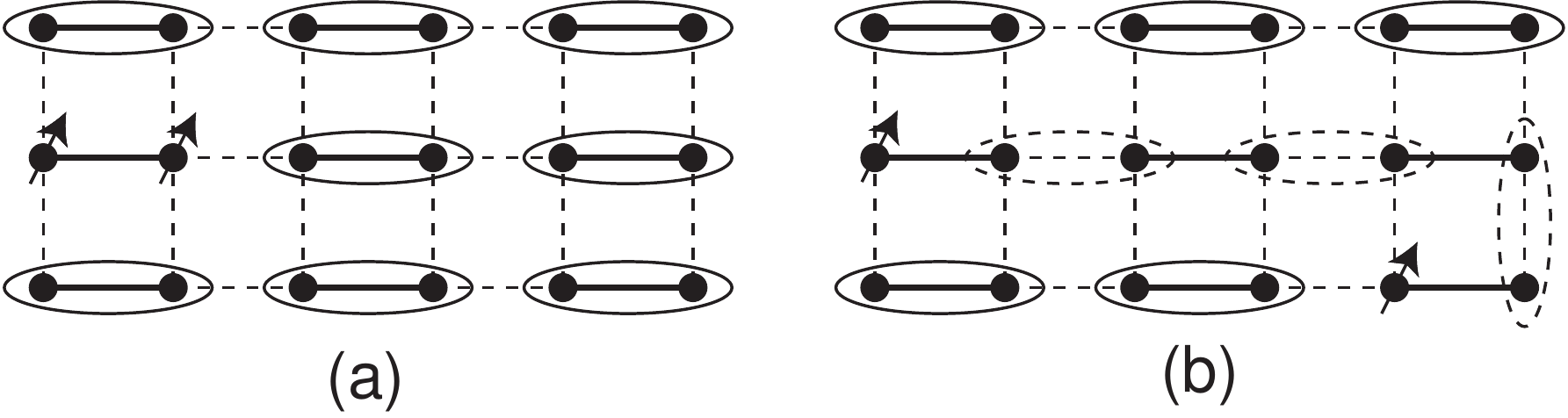}
\caption{(a) Cartoon picture of the bosonic $S=1$ excitation of
the paramagnet. (b) Fission of the $S=1$ excitation into two
$S=1/2$ spinons. The spinons are connected by a ``string'' of
valence bonds (denoted by dashed ovals) which lie on weaker bonds;
this string costs a finite energy per unit length and leads to the
confinement of spinons.} \label{fig3}
\end{figure}

Let us turn now to the vicinity of $g = 0$. Exactly at
$g=0$, $H_d$ is the Hamiltonian of a set of decoupled
dimers, with the simple exact ground state wavefunction shown in
Fig.~\ref{fig2}: the spins in each dimer pair into valence bond
singlets, leading to a paramagnetic state which preserves spin
rotation invariance and all lattice symmetries.
Excitations are now formed by breaking a valence bond, which leads
to a {\em three}-fold degenerate state with total spin $S=1$, as
shown in Fig.~\ref{fig3}a. At $g=0$, this broken bond is
localized, but at finite $g$ it can hop from site-to-site,
leading to a triplet quasiparticle excitation.
Note that this quasiparticle is {\em not\/} a spin-wave (or
equivalently, a `magnon') but is more properly referred to as a
spin 1 {\em exciton} or a {\em triplon}. We
parameterize its energy at small wavevectors $k$ (measured from
the minimum of the spectrum in the Brillouin zone) by
\begin{equation}
\varepsilon_k = \Delta + \frac{c_x^2 k_x^2 + c_y^2 k_y^2}{2
\Delta}, \label{epart}
\end{equation}
where $\Delta$ is the spin gap, and $c_x$, $c_y$ are velocities;
we will provide an explicit derivation of (\ref{epart}) in
Section~\ref{sec:lad2}. Fig.~\ref{fig3} also presents a simple
argument which shows that the $S=1$ exciton cannot fission into
two $S=1/2$ `spinons'.

\begin{figure}[t]
\centering
\includegraphics[width=4.5in]{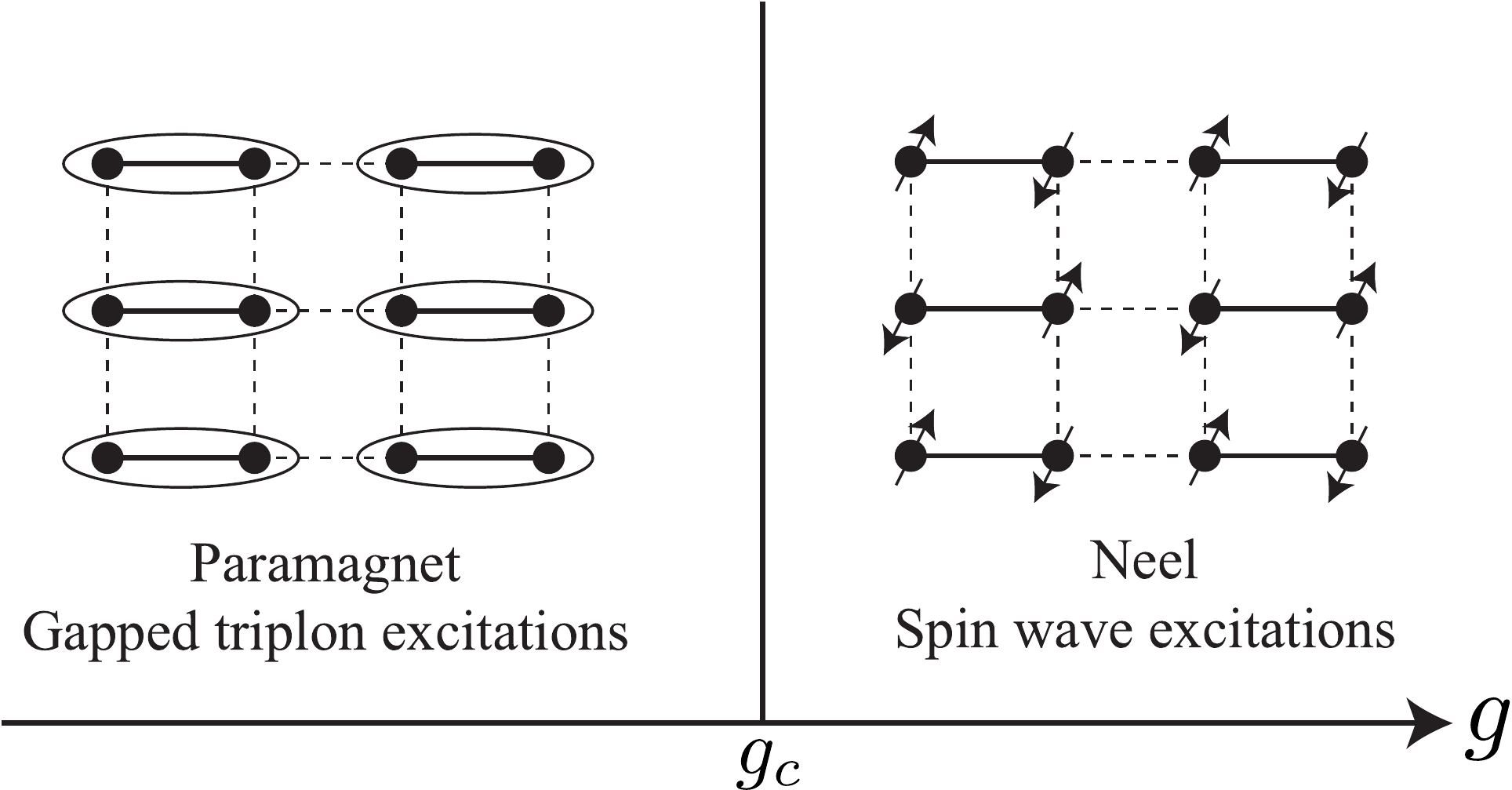}
\caption{Ground states of $H_d$ as a function of $g$. The
quantum critical point is at \protect\cite{matsumoto} $g_c =
0.52337(3)$. The compound TlCuCl$_3$ undergoes a similar quantum
phase transition under applied pressure \cite{oosawa,ssruegg}.
}\label{fig4}
\end{figure}

The very distinct symmetry signatures of the ground states and
excitations between $g \approx 1$ and $g \approx 0$
make it clear that the two limits cannot be continuously
connected. It is known that there is an intermediate second-order
phase transition at \cite{gsh,matsumoto} $g = g_c =
0.52337(3)$ between these states as shown in Fig.~\ref{fig4}.
Both the spin gap $\Delta$ and the N\'{e}el order parameter $N_0$
vanish continuously as $g_c$ is approached from either side.

\subsubsection{Bond operators and quantum field theory}
\label{sec:lad2}

In this section we will develop a continuum description of the low
energy excitations in the vicinity of the critical point
postulated above. There are a number of ways to obtain the same
final theory: here we will use the method of {\em bond operators}
\cite{bondops,chubukov}, which has the advantage of making the
connection to the lattice degrees of freedom most direct. We
rewrite the Hamiltonian using bosonic operators which reside on
the centers of the $\mathcal{A}$ links so that it is explicitly
diagonal at $g=0$. There are 4 states on each $\mathcal{A}$
link ($\left| \uparrow \uparrow \right\rangle$, $\left| \uparrow
\downarrow \right\rangle$, $\left| \downarrow \uparrow
\right\rangle$, and $\left| \downarrow \downarrow \right\rangle$)
and we associate these with the canonical singlet boson $s$ and
the canonical triplet bosons $t_{a}$ ($a=x,y,z$) so that
\begin{eqnarray}
|s \rangle \equiv  s^{\dagger} | 0 \rangle  =  \frac{1}{\sqrt{2}}
\left( |\uparrow\downarrow\rangle -
|\downarrow\uparrow\rangle\right) ~~;~~ |t_x \rangle \equiv
t_{x}^{\dagger} | 0 \rangle & = & \frac{-1}{\sqrt{2}} \left(
|\uparrow\uparrow\rangle - |\downarrow\downarrow\rangle\right)~~;
\nonumber \\
|t_y \rangle \equiv t_{y}^{\dagger} | 0 \rangle  =
\frac{i}{\sqrt{2}} \left( |\uparrow\uparrow\rangle +
|\downarrow\downarrow\rangle\right) ~~;~~ | t_z \rangle \equiv
t_{z}^{\dagger} | 0 \rangle & = & \frac{1}{\sqrt{2}} \left(
|\uparrow\downarrow\rangle + |\downarrow\uparrow\rangle\right).
\label{states}
\end{eqnarray}
Here $|0 \rangle$ is some reference vacuum state which does not
correspond to a physical state of the spin system. The physical
states always have a single bond boson and so satisfy the
constraint
\begin{equation}
s^{\dagger} s + t^{\dagger}_{a} t_{a} = 1 \, .
\label{const}
\end{equation}
By considering the various matrix elements $\langle s | \vec{S}_1
| t_{a} \rangle$, $\langle s | \vec{S}_2 | t_{a}
\rangle$, $\ldots$, of the spin operators $\vec{S}_{1,2}$ on the
ends of the link, it follows that the action of $\vec{S}_1$ and
$\vec{S}_2$ on the singlet and triplet states is equivalent to the
operator identities
\begin{eqnarray}
S_{1a} & = & \frac{1}{2}\left( s^{\dagger} t_{a} +
t_{a}^{\dagger} s - i\epsilon_{a  b  c}
t_{  b}^{\dagger} t_{  c} \right), \nonumber  \\
S_{2a} & = & \frac{1}{2}\left( - s^{\dagger} t_{a} -
t_{a}^{\dagger} s - i\epsilon_{a  b  c}
t_{  b}^{\dagger} t_{  c} \right). \label{rep2}
\end{eqnarray}
where $a$,$  b$,$  c$ take the values $x$,$y$,$z$,
repeated indices are summed over and $\epsilon$ is the totally
antisymmetric tensor. Inserting (\ref{rep2}) into (\ref{ham}), and
using (\ref{const}), we find the following Hamiltonian for the
bond bosons:
\begin{eqnarray}
H_{d} &=& H_0 + H_1 \nonumber \\
H_0 &=& J \sum_{\ell \in \mathcal{A}} \left( - \frac{3}{4}
s_{\ell}^{\dagger} s_{\ell} + \frac{1}{4} t_{\ell a}^{\dagger}
t_{\ell a}
\right) \nonumber \\
H_1 &=& g J \sum_{\ell,m \in \mathcal{A}} \Biggl[ a(\ell,m)
\left( t_{\ell  a}^{\dagger}  t_{m  a} s_{m}^{\dagger}
s_{\ell} + t_{\ell  a}^{\dagger}  t_{m  a}^{\dagger} s_{m}
s_{\ell} +
{\rm H.c.} \right) + b(\ell,m ) \nonumber \\
&~& \!\!\!\!\!\!\!\!\!\!\!\!\!\!\!\!\!\!\!\!\!\times \left( i
\epsilon_{ a  b  c} t_{m  a}^{\dagger} t_{\ell
  b}^{\dagger} t_{\ell   c} s_{m} + {\rm H.c.} \right) +
c(\ell,m ) \left( t_{\ell  a}^{\dagger} t_{m a}^{\dagger}
t_{m  b} t_{\ell  b} - t_{\ell  a}^{\dagger}  t_{m
  b}^{\dagger} t_{m a} t_{\ell   b} \right) \Biggr] \, ,
\label{bondham}
\end{eqnarray}
where $\ell,m$ label links in $\mathcal{A}$, and $a,b,c$ are
numbers associated with the lattice couplings which we will not
write out explicitly. Note that $H_1 =0 $ at $g=0$, and so
the spectrum of the paramagnetic state is fully and exactly
determined. The main advantage of the present approach is that
application of the standard methods of many body theory to
(\ref{bondham}), while imposing the constraint (\ref{const}),
gives a very satisfactory description of the phases with $g
\neq 0$, including across the transition to the N\'{e}el state. In
particular, an important feature of the bond operator approach is
that the simplest mean field theory already yields ground states
and excitations with the correct quantum numbers; so a strong
fluctuation analysis is not needed to capture the proper physics.

A complete numerical analysis of the properties of (\ref{bondham})
in a self-consistent Hartree-Fock treatment of the four boson
terms in $H_1$ has been presented in Ref.~\cite{bondops}. In all
phases the $s$ boson is well condensed at zero momentum, and the
important physics can be easily understood by examining the
structure of the low-energy action for the $t_{ a}$ bosons.
For the particular Hamiltonian (\ref{ham}), the spectrum of the
$t_{ a}$ bosons has a minimum at the momentum $(0,\pi)$, and
for large enough $g$ the $t_{ a}$ condense at this
wavevector: the representation (\ref{rep2}) shows that this
condensed state is the expected N\'{e}el state, with the magnetic
moment oscillating as in (\ref{neel}). The condensation transition
of the $t_{ a}$ is therefore the quantum phase transition
between the paramagnetic and N\'{e}el phases of the coupled dimer
antiferromagnet. In the vicinity of this critical point, we can
expand the $t_{ a}$ bose field in gradients away from the
$(0,\pi)$ wavevector: so we parameterize
\begin{equation}
t_{\ell,  a} (\tau) = t_{ a} (r_{\ell}, \tau) e^{i (0,\pi)
\cdot \vec r_{\ell}} \label{t1}
\end{equation}
where $\tau$ is imaginary time, $\vec{r} \equiv (x,y)$ is a continuum
spatial coordinate, and expand the effective action in spatial
gradients. In this manner we obtain
\begin{eqnarray}
\mathcal{S}_t &=& \int d^2 r d \tau \left[ t^{\dagger}_{ a}
\frac{\partial t_{ a}}{\partial \tau} + C t^{\dagger}_{ a}
t_{ a} - \frac{D}{2} \left( t_{ a} t_{ a} + {\rm H.c.}
\right) + K_{1x} |\partial_x t_{ a} |^2 +
K_{1y} |\partial_y t_{ a} |^2\right. \nonumber \\
&~&~~~~~~~~~~~~~~~~~~~~\left. + \frac{1}{2} \left( K_{2x}
(\partial_x t_{ a} )^2 + K_{2y} (\partial_y t_{ a} )^2 +
{\rm H.c.} \right) + \cdots \right]. \label{st}
\end{eqnarray}
Here $C,D,K_{1,2 x,y}$ are constants that are determined by the
solution of the self-consistent equations, and the ellipses
represent terms quartic in the $t_{ a}$. The action
$\mathcal{S}_t$ can be easily diagonalized, and we obtain a $S=1$
quasiparticle excitation with the spectrum
\begin{equation}
\varepsilon_k = \left[ \left(C + K_{1x} k_x^2 + K_{1y} k_y^2
\right)^2 - \left( D + K_{2x} k_x^2 + K_{2y} k_y^2 \right)^2
\right]^{1/2}. \label{epart2}
\end{equation}
This is, of course, the triplon (or spin exciton) excitation of
the paramagnetic phase postulated earlier in (\ref{epart}); the
latter result is obtained by expanding (\ref{epart2}) in momenta,
with $\Delta = \sqrt{C^2 - D^2}$. This value of $\Delta$ shows
that the ground state is paramagnetic as long as $C>D$, and the
quantum critical point to the N\'{e}el state is at $C=D$.

The critical point and the N\'{e}el state are more conveniently
described by an alternative formulation of $\mathcal{S}_t$
(although an analysis using bond operators directly is also
possible \cite{sommer}). It is useful to decompose the complex
field $t_{ a}$ into its real and imaginary parts as follows
\begin{equation}
t_{ a} = Z (\varphi_{ a} + i \pi_{ a} ), \label{tphi}
\end{equation}
where $Z$ is a normalization chosen below.
From (\ref{t1}) and the connection to the lattice spin operators,
it is not difficult to show that the vector ${\varphi}_a$ is proportional
to the N\'eel order parameter $\vec{n}$ in Eq.~(\ref{neel}).
Insertion of
(\ref{tphi}) into (\ref{st}) shows that the field $\pi_{ a}$
has a quadratic term $\sim (C+D) \pi_{ a}^2$, and so the
coefficient of $\pi_{ a}^2$ remains large even as the spin
gap $\Delta$ becomes small. Consequently, we can safely integrate
$\pi_{ a}$ out, and the resulting action for
the N\'eel order parameter ${\varphi}_a$ takes the form
\begin{equation}
\mathcal{S}_{\varphi} = \int d^2 r d \tau \left[ \frac{1}{2}
\left\{ \left(
\partial_{\tau} \varphi_{ a} \right)^2 + c_x^2 \left(
\partial_{x} \varphi_{ a} \right)^2 + c_y^2 \left(
\partial_{y} \varphi_{ a} \right)^2 + s \varphi_{ a}^2
\right\} + \frac{u}{24} \left( \varphi_{ a}^2 \right)^2
\right]. \label{sp}
\end{equation}
Here we have chosen $Z$ to fix the coefficient of the temporal
gradient term, and $s = C^2 - D^2$.

\begin{figure}[t]
\begin{center}
 \includegraphics[width=3in]{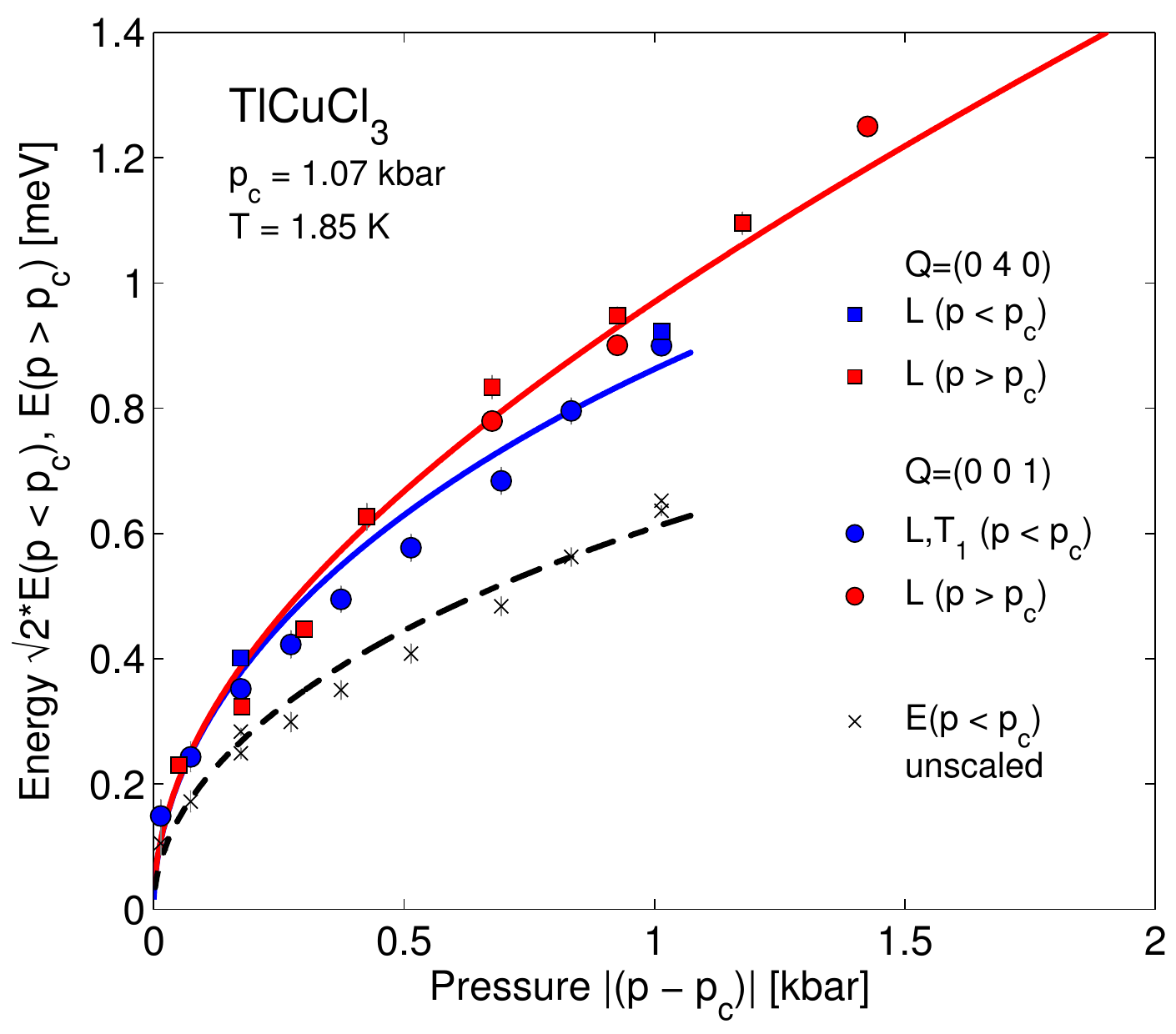}
 \caption{Energies of the gapped collective modes across the pressure ($p$) tuned quantum phase transition in
 TlCuCl$_3$ observed by Ruegg {\em et al.}\cite{ssruegg}. We test the description by the action $\mathcal{S}_{\varphi}$
 in Eq.~(\ref{sp}) with $s \propto (p_c - p)$ by comparing $\sqrt{2}$ times the energy gap for $p<p_c$ with the
 energy of the longitudinal mode for $p>p_c$. The lines are the fits to a $\sqrt{|p-p_c|}$ dependence, testing the 1/2
 exponent.}
\label{fig:ruegg}
\end{center}
\end{figure}

The action $\mathcal{S}_\varphi$ gives a simple picture of excitations
across the quantum critical point, which can be quantitatively compared to neutron scattering experiments \cite{ssruegg}
on TlCuCl$_3$.
In the paramagnetic phase
($s>0$), a triplet of gapped excitations is observed, corresponding to the three normal modes of ${\varphi}_ a$ oscillating
about $\varphi_ a = 0$; as expected, this triplet gap vanishes upon approaching the quantum critical point. In a mean field analysis,
the field theory in Eq.~(\ref{sp}) has a triplet gap of $\sqrt{s}$ (mean field theory is applicable to TlCuCl$_3$ because
this antiferromagnet is three dimensional).
In the N\'eel phase, the neutron scattering detects 2 gapless spin waves, and one gapped longitudinal
mode \cite{normand} (the gap to this longitudinal mode vanishes at the quantum critical point), as is expected from fluctuations
in the inverted `Mexican hat' potential of $\mathcal{S}_{\varphi}$ for $s<0$. The longitudinal mode has a mean-field
energy gap of $\sqrt{2 |s|}$.
These mean field predictions for the energy of the gapped modes on the two sides of the transition are tested in Fig.~\ref{fig:ruegg}:
the observations are in good agreement with the 1/2 exponent and the
predicted \cite{sslg,solvay} $\sqrt{2}$ ratio, providing a non-trival experimental
test of the $\mathcal{S}_{\varphi}$ field theory.

We close this subsection by noting that all of the above results
have a direct generalization to other lattices. One important
difference that emerges in such calculations on some frustrated
lattices \cite{balents} is worth noting explicitly here: the
minimum of the $t_{ a}$ spectrum need not be at special
wavevector like $(0,\pi)$, but can be at a more generic wavevector
$\vec{K}$ such that $\vec{K}$ and $-\vec{K}$ are not separated by
a reciprocal lattice vector. A simple example which we consider
here is an extension of (\ref{ham}) in which there are additional
exchange interactions along all diagonal bonds oriented
`north-east' (so that the lattice has the connectivity of a
triangular lattice). In such cases, the structure of the low
energy action is different, as is the nature of the magnetically
ordered state. The parameterization (\ref{t1}) must be replaced by
\begin{equation}
t_{\ell  a} (\tau) = t_{1  a} (r_{\ell}, \tau) e^{i \vec K
\cdot \vec r_{\ell}} + t_{2  a} (r_{\ell}, \tau) e^{-i \vec K
\cdot \vec r_{\ell}} \, , \label{t2}
\end{equation}
where $t_{1,2 a}$ are independent complex fields. Proceeding
as above, we find that the low-energy effective action (\ref{sp})
is replaced by
\begin{eqnarray}
\mathcal{S}_{\Phi} &=& \int d^2 r d \tau \biggl[ \left|
\partial_{\tau} \Phi_{ a} \right|^2 + c_x^2 \left|
\partial_{x} \Phi_{ a} \right|^2 + c_y^2 \left|
\partial_{y} \Phi_{ a} \right|^2 + s
\left|\Phi_{ a}\right|^2 \nonumber \\
&~&~~~~~~~~~~~~~~~~~~~~~~~~~~ +\frac{u}{2} \left(
\left|\Phi_{ a} \right|^2 \right)^2 + \frac{v}{2} \left|
\Phi_{ a}^2 \right|^2 \biggr]. \label{sp2}
\end{eqnarray}
where now $\Phi_{ a}$ is a {\em complex} field such that
$\langle \Phi_{ a} \rangle \sim \langle t_{1 a} \rangle
\sim \langle t^{\dagger}_{2 a} \rangle$. Notice that there is
now a second quartic term with coefficient $v$. If $v> 0$,
configurations with $\Phi_{ a}^2 = 0$ are preferred: in such
configurations $\Phi_{ a} = n_{1 a} + i n_{2  a}$,
where $n_{1,2 a}$ are two equal-length orthogonal vectors.
Then from (\ref{t2}) and (\ref{rep2}) it is easy to see that the
physical spins possess {\em spiral} order in the magnetically
ordered state in which $\Phi_{ a}$ is condensed. For the case
$v < 0$, the optimum configuration has $\Phi_{ a} = n_{ a}
e^{i \theta}$ where $n_{ a}$ is a real vector: this leads to a
magnetically ordered state with spins polarized {\em collinearly}
in a spin density wave at the wavevector $\vec{K}$. The critical properties of the
model in Eq.~(\ref{sp2}) have been described in Ref.~\cite{vicari0}.

\subsection{Frustrated square lattice antiferromagnets: Schwinger bosons}
\label{sec:frust}

As discussed at the beginning of Section~\ref{sec:insulator}, the more important
and complex cases of quantum antiferromagnets are associated with those
that have a single $S=1/2$ spin per unit cell.
Such models are more likely to have phases in which the exotic spinon excitations
of Fig.~\ref{fig3} are deconfined, {\em i.e.\/}, their ground states possess
neutral $S=1/2$ excitations and `topological' order. We will meet the earliest established
examples \cite{self4,wen1} of such phases below.

\begin{figure}[t]
\centering
\includegraphics[width=2in]{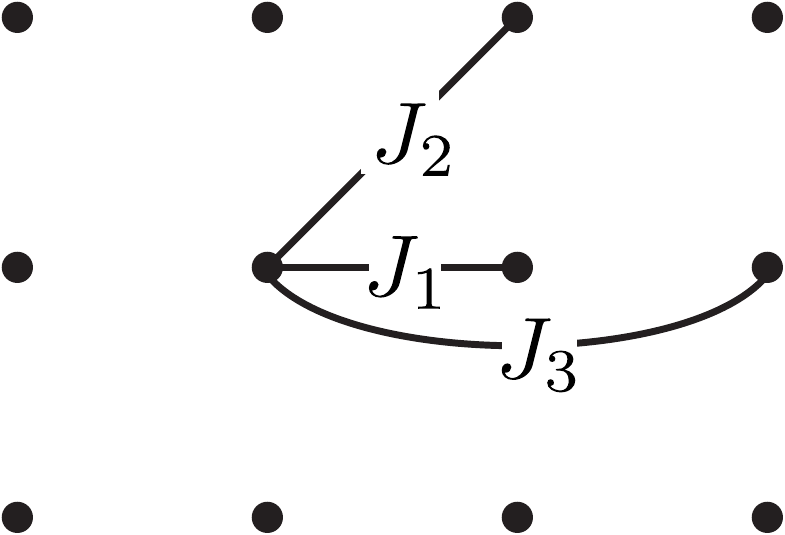}
\caption{The $J_1$-$J_2$-$J_3$ antiferromagnet. Spin $S$ spins are placed on each
site of the square lattice, and they are coupled to all first, second, and third neighbors as shown.
The Hamiltonian has the full space group symmetry of the square lattice, and there is only
one spin per unit cell.}
\label{fig:j1j2j3}
\end{figure}

We are interested in Hamiltonians of the form
\begin{equation}
{\cal H} = \sum_{i,j} J_{ij} \vec{S}_i \cdot \vec{S}_j
\label{hamil}
\end{equation}
where we consider the general case of $\vec{S}_i$ being spin $S$
quantum spin operators on the sites, $i$, of
a $2$-dimensional lattice. The $J_{ij}$ are short-ranged antiferromagnetic
exchange interactions. We will mainly consider here the so-called
square lattice $J_1$-$J_2$-$J_3$ model, which has first, second, and third neighbor
interactions (see Fig.~\ref{fig:j1j2j3}).
Similar results have also been obtained on the triangular and kagome lattices \cite{self6,wang}.

The main direct applications of the results here are to experiments on a variety
of two-dimensional Mott insulators on the square, triangular, and kagome lattices.
As noted earlier, we direct the reader to Ref.~\cite{solvay} for a discussion
of these experiments. There have also
been extensive numerical studies, also reviewed in the previous article \cite{solvay}, which
are in good accord with the phase diagrams presented below. Applications to the cuprates, and
to Fig.~\ref{fig:figglobal}, will be discussed
in the following sections.

A careful examination of the non-magnetic `spin-liquid' phases requires an
approach which is designed explicitly to be valid in a region well
separated from N\'eel long range order, and preserves SU(2) symmetry
at all stages. It should also be designed to naturally allow for neutral $S=1/2$
excitations. To this end, we introduce the Schwinger
boson description \cite{assa}, in terms of elementary $S=1/2$ bosons.
For the group $SU(2)$ the complete set of $(2S+1)$
states on site $i$ are represented as follows
\begin{equation}
|S , m \rangle \equiv \frac{1}{\sqrt{(S+m)! (S-m)!}}
(b_{i\uparrow}^{\dagger})^{S+m}
(b_{i\downarrow}^{\dagger})^{S-m} | 0 \rangle,
\end{equation}
where $m = -S, \ldots S$ is the $z$ component of the spin ($2m$ is an integer).
We have introduced two flavors of bosons on each site,
created by the canonical operator
$b_{i\alpha}^{\dagger}$, with $\alpha = \uparrow, \downarrow$, and
$|0\rangle$ is the vacuum with no bosons. The total number of bosons, $n_b$
is the same for all the states; therefore
\begin{equation}
b_{i\alpha}^{\dagger}b_{i}^{\alpha} = n_b
\label{boseconst}
\end{equation}
with $n_b = 2S$ (we will henceforth assume an implied summation over
repeated upper and lower indices). It is not difficult to see that the above
representation of the states is completely equivalent to the following
operator identity between the spin and boson operators
\begin{equation}
{S}_{ia} =  \frac{1}{2}
b_{i\alpha}^{\dagger} {\sigma^{a\alpha}}_{\beta} b_{i}^{\beta} \, ,
\end{equation}
where $a=x,y,z$ and the $\sigma^a$ are the usual $2\times 2$ Pauli
matrices.
The spin-states on two sites $i,j$ can combine to form a singlet in a unique
manner - the wavefunction of the singlet state is particularly simple
in the boson formulation:
\begin{equation}
\left( \varepsilon^{\alpha\beta} b_{i\alpha}^{\dagger}
b_{j\beta}^{\dagger} \right)^{2S} |0\rangle \, .
\end{equation}
Finally we note that, using the constraint (\ref{boseconst}), the
following Fierz-type identity can be established
\begin{equation}
\left( \varepsilon^{\alpha \beta}
b_{i \alpha}^{\dagger} b_{j \beta}^{\dagger} \right)
\left( \varepsilon_{\gamma \delta}
b_{i}^{\gamma} b_{j}^{\delta} \right) =
 - 2 \vec{S}_i \cdot \vec{S}_j
+ n_b^2 /2 + \delta_{ij} n_b
\label{su2}
\end{equation}
where $\varepsilon$ is the totally antisymmetric $2\times2$ tensor
\begin{equation}
\varepsilon = \left( \begin{array}{cc}
0 & 1 \\
-1 & 0 \end{array} \right) \, .
\end{equation}
This implies that ${\cal H}$ can be rewritten in the form (apart from
an additive constant)
\begin{equation}
{\cal H} = - \frac{1}{2} \sum_{\langle ij\rangle} J_{ij} \left( \varepsilon^{\alpha \beta}
b_{i \alpha}^{\dagger} b_{j \beta}^{\dagger} \right)
\left( \varepsilon_{\gamma \delta}
b_{i}^{\gamma} b_{j}^{\delta} \right) \, .
\label{hafkag}
\end{equation}
This form makes it clear that ${\cal H}$ counts the number of singlet
bonds.

We have so far defined a one-parameter ($n_b$) family of models ${\cal
H}$ for a fixed realization of the $J_{ij}$. Increasing $n_b$ makes
the system more classical and a large $n_b$ expansion is therefore not
suitable for studying the quantum-disordered phase. For this reason we
introduce a second parameter - the flavor index $\alpha$ on the bosons
is allowed to run from $1 \ldots 2N$ with $N$ an arbitrary integer.
This therefore allows the bosons to transform under $SU(2N)$
rotations.
However the $SU(2N)$ symmetry turns out to be too large. We want to
impose the additional restriction that the spins on a pair of sites be
able to combine to form a singlet state, thus generalizing the
valence-bond structure of $SU(2)$ - this valence-bond formation is
clearly a crucial feature determining the structure of the quantum
disordered phase. It is well-known that this is
impossible for $SU(2N)$ for $N>1$ - there is no generalization of the
second-rank, antisymmetric, invariant
tensor $\varepsilon$ to general $SU(2N)$.

The proper generalization turns out to be to the group
$Sp(N)$~\cite{self4}. This
group is defined by the set
of $2N\times2N$ unitary matrices $U$
such that
\begin{equation}
U^{T} \mathcal{J} U = \mathcal{J}
\label{dsp2}
\end{equation}
where
\begin{equation}
\mathcal{J}_{\alpha\beta} = \mathcal{J}^{\alpha\beta} =
\LEFTRIGHT(){\begin{array}{cccccc}
  & 1 & & & & \\
-1 &  & & & & \\
 & &  & 1  & & \\
 & & -1 &  & & \\
 & &  & & \ddots & \\
 & & & & & \ddots
\end{array}}
\end{equation}
is the generalization of the $\varepsilon$ tensor to $N>1$; it has $N$
copies of $\varepsilon$ along the diagonal. It is
clear that $Sp(N) \subset SU(2N)$ for $N>1$, while $Sp(1) \cong SU(2)$.
The $b_{i}^{\alpha}$ bosons transform as the fundamental
representation of $Sp(N)$; the ``spins'' on the lattice
therefore belong to the symmetric product of $n_b$
fundamentals, which is also an irreducible representation.
Valence bonds
\begin{equation}
\mathcal{J}^{\alpha\beta} b_{i\alpha}^{\dagger}
b_{j\alpha}^{\dagger}
\end{equation}
can be formed between any two sites; this operator is a
singlet under $Sp(N)$ because of (\ref{dsp2}). The form (\ref{hafkag})
of ${\cal H}$ has a natural generalization to general $Sp(N)$:
\begin{equation}
{\cal H}= -\sum_{i>j} \frac{J_{ij}}{2N} \left(
\mathcal{J}^{\alpha\beta}
b_{i\alpha}^{\dagger} b_{j,\beta}^{\dagger} \right)
\left( \mathcal{J}_{\gamma\delta} b_{i}^{\gamma} b_{j}^{\delta}
\right)
\label{hex}
\end{equation}
where the indices $\alpha,\beta,\gamma,\delta$ now run over $1\ldots
2N$.
We recall also that the constraint (\ref{boseconst}) must be imposed
on every site of the lattice.

\begin{figure}[t]
\centering
\includegraphics[width=4.5in]{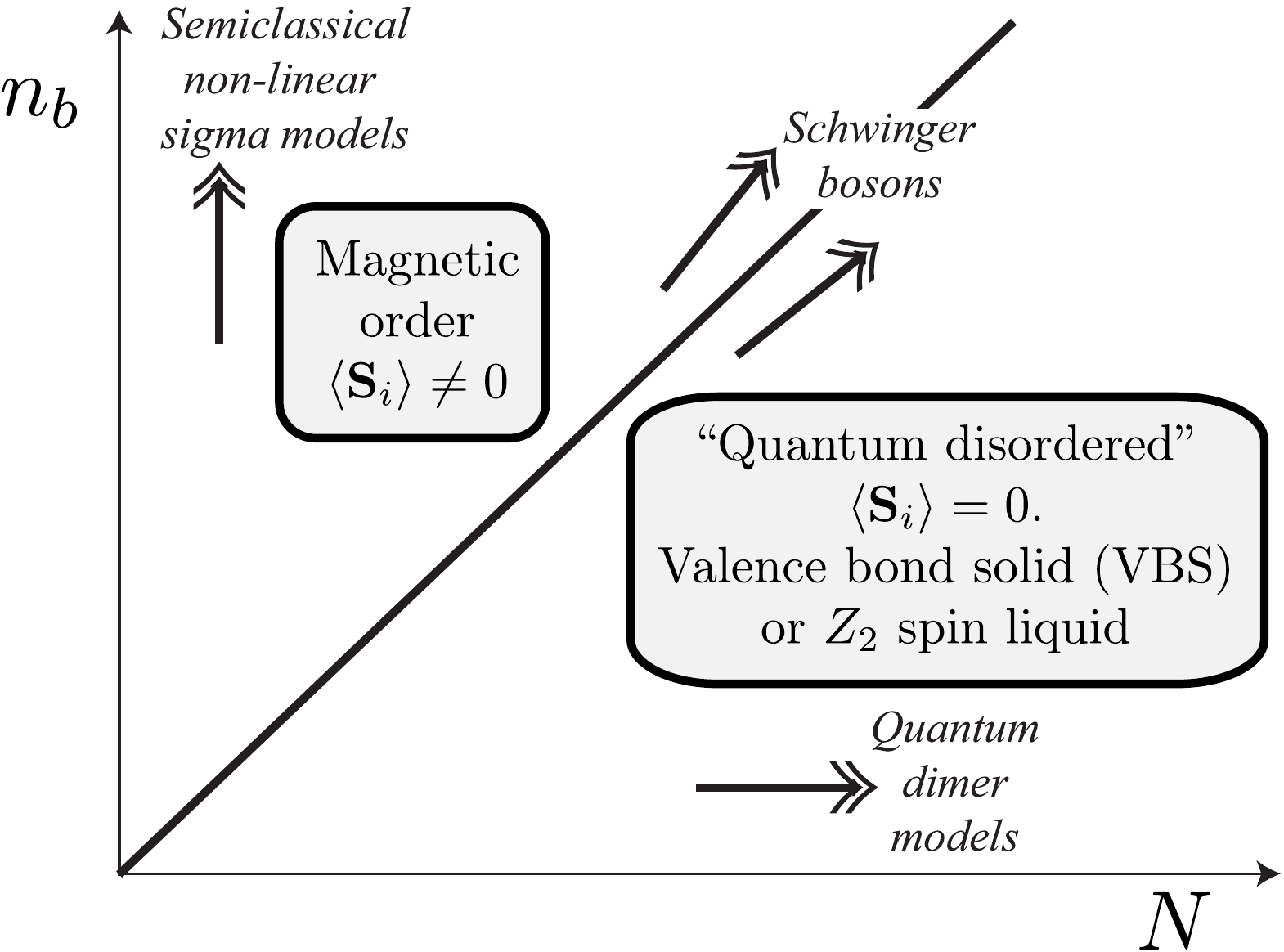}
\caption{Phase diagram of the 2D $Sp(N)$ antiferromagnet
${\cal H}$ as a function of the ``spin'' $n_b$; from Refs.~\cite{self1,self4,self2,self3,self5} The ``quantum disordered'' region
preserves Sp($N$) spin rotation invariance, and there is no magnetic long-range order;
however, the ground states here have
new types of emergent order (VBS or $Z_2$ topological order),
which are described in the text. On the square lattice, the $Z_2$ spin liquid phases also break a global
lattice rotational symmetry, and so they have `Ising-nematic' order; the $Z_2$ spin liquids on the
triangular and kagome lattices do not break any lattice symmetry.}
\label{phasediag}
\end{figure}

We now have a two-parameter ($n_b , N$) family of models ${\cal H}$
for a fixed realization of the $J_{ij}$. It is very instructive to
consider the phase diagram of ${\cal H}$ as a function of these two
parameters (Fig.~\ref{phasediag}).

The limit of large $n_b$, with $N$ fixed leads to the semi-classical
theory. For the special case of $SU(2)$ antiferromagnets with a
two-sublattice collinear N\'eel ground state, the semiclassical
fluctuations are described by the $O(3)$ non-linear sigma model.
For other
models~\cite{self1,ian1,johan,premi,halpsas,dombread,joli},
the structure of the non-linear sigma
models is rather more complicated and will not be considered here.

A second limit in which the problem simplifies is $N$
large at fixed $n_b$~\cite{brad,self1}.
It can be shown that in this limit the ground
state is quantum disordered. Further, the
low-energy dynamics of ${\cal H}$ is described by an effective quantum-dimer
model~\cite{dan,self1}, with
each dimer configuration representing a particular pairing
of the sites into valence-bonds.
There have been extensive studies of such quantum dimer models
which we will not review here. All the quantum dimer model
studies in the ``quantum disordered'' region
of Fig.~\ref{phasediag} have yielded phases which were obtained earlier \cite{self4}
by the methods to be described below.

The most interesting solvable limit is obtained by fixing the ratio of
$n_b$ and $N$
\begin{equation}
\kappa = \frac{n_b}{N}
\end{equation}
and subsequently taking the limit of large $N$~\cite{assa}; this limit will be
studied in this section in considerable detail. The implementation of
${\cal H}$ in terms of bosonic operators also turns out to be
naturally suited for studying this limit. The parameter $\kappa$
is arbitrary; tuning $\kappa$ modifies the slope of the line in
Fig.~\ref{phasediag} along which the large $N$ limit is taken. From
the previous limits discussed above, one might expect that the ground
state of ${\cal H}$ has magnetic long range order (LRO) for large $\kappa$ and is
quantum-disordered for small $\kappa$.
We will indeed find below that for any set of $J_{ij}$ there is a
critical value of $\kappa = \kappa_c$ which separates the
magnetically ordered and the quantum disordered phase.

The transition at $\kappa=\kappa_c$ is second-order
at $N=\infty$, and is a powerful feature of the present large-$N$ limit. In the
vicinity of the phase transition, we expect the physics to be
controlled by long-wavelength, low-energy spin fluctuations; the
large-$N$ method offers an unbiased guide in identifying the proper
low-energy degress of freedom and determines the effective action
controlling them. Having obtained a long-wavelength continuum theory
near the transition, one might hope to analyze the continuum theory
independently of the large-$N$ approximation and obtain results that
are more generally valid.

We will discuss the structure of the $N=\infty$ mean-field theory ,
with $n_b = \kappa N$ in
Section~\ref{sec:6A}. The long-wavelength effective actions will be derived and
used to describe general properties of the phases and the phase transitions in
Section~\ref{sec:6B}.

\subsubsection{Mean-field theory}
\label{sec:6A}

We begin by analyzing ${\cal H}$ at $N = \infty$
with $n_b = \kappa N$. As noted above, this limit is most conveniently
taken using the bosonic operators.
We may represent the partition function of ${\cal H}$ by
\begin{equation}
Z = \int {\cal D} Q {\cal D} b  {\cal D} \lambda
\exp \left( - \int_0^{\beta} {\cal L} d\tau \right) ,
\end{equation}
where
\begin{displaymath}
{\cal L} = \sum_{i} \left [
b_{i\alpha }^{\dagger}  \left( \frac{d}{d\tau} + i\lambda_i \right)
b_i^{\alpha} - i\lambda_i n_b \right ]~~~~~~~~~~~~~~~~~~~~~~
\end{displaymath}
\begin{equation}
~~~~~~~~~~~+  \sum_{\langle i,j\rangle} \left [
N \frac{J_{ij} |Q_{i,j} |^2}{2}
- \frac{J_{ij} Q_{i,j}^{\ast}}{2} \mathcal{J}_{\alpha\beta} b_i^{\alpha}
b_j^{\beta}
+ H.c. \right] .
\label{zfunct}
\end{equation}
Here the $\lambda_i$
fix the boson number of $n_b$ at each site;
$\tau$-dependence of all fields is implicit. The complex field $Q$ was introduced by
a Hubbard-Stratonovich decoupling of ${\cal H}$: performing the functional integral
over $Q$ reproduces the exchange coupling in Eq.~(\ref{hex}).
An important feature
of the lagrangian
${\cal L}$ is its $U(1)$ gauge invariance under which
\begin{eqnarray}
b_{i\alpha}^{\dagger} & \rightarrow & b_{i\alpha}^{\dagger} (i)
\exp \left( i\rho_i (\tau ) \right ) \nonumber \\
Q_{i,j} &\rightarrow & Q_{i,j} \exp \left( - i \rho_i (\tau ) - i
\rho_j (\tau ) \right) \nonumber \\
\lambda_i & \rightarrow & \lambda_i + \frac{\partial \rho_i}{\partial
\tau} (\tau) .
\label{gaugetrans}
\end{eqnarray}
The functional integral over ${\cal L}$
faithfully represents the partition function apart from an overall factor associated 
with this gauge redundancy.

The $1/N$ expansion of the free energy can be obtained by integrating out
of ${\cal L}$ the $2N$-component $b$,$\bar{b}$ fields to leave an effective action
for $Q$, $\lambda$ having coefficient $N$ (because $n_b \propto N$).
Thus the $N \rightarrow \infty$
limit is given by minimizing the effective action with respect to
``mean-field'' values of $Q = \bar{Q}$, $i \lambda = \bar{\lambda}$ (we are
ignoring here the possibility of magnetic LRO which requires an
additional condensate $x^{\alpha} = \langle b^{\alpha} \rangle$ - this
has been discussed elsewhere~\cite{self4,self5}).
This is in turn equivalent to solving
the mean-field Hamiltonian
\begin{eqnarray}
{\cal H}_{MF} &=&  \sum_{\langle i,j\rangle} \left(
N \frac{J_{ij} |\bar{Q}_{ij} |^2}{2}
- \frac{J_{ij} \bar{Q}_{i,j}^{\ast}}{2} \mathcal{J}_{\alpha\beta} b_i^{\alpha}
b_j^{\beta}
+ H.c. \right) \nonumber \\
&& + \sum_{i} \bar{\lambda}_i (
b_{i\alpha }^{\dagger} b_{i}^{\alpha} -  n_b ) \, .
\end{eqnarray}
This Hamiltonian is quadratic in the boson operators and all its
eigenvalues can be determined by a Bogoliubov transformation. This
leads in general to an expression of the form
\begin{equation}
{\cal H}_{MF} = E_{MF}[ \bar{Q} , \bar{\lambda}] + \sum_{\mu} \omega_{\mu} [\bar{Q} , \bar{\lambda}]
\gamma_{\mu\alpha}^{\dagger} \gamma_{\mu}^{\alpha} \, .
\end{equation}
The index $\mu$ extends over $1\ldots$ number of sites in the system,
$E_{MF}$ is the ground state energy and is a functional of $\bar{Q}$,
$\bar{\lambda}$, $\omega_{\mu}$ is the eigenspectrum of excitation energies
which is a also a function of $\bar{Q}$, $\bar{\lambda}$, and the
$\gamma_{\mu}^{\alpha}$ represent the bosonic eigenoperators. The
excitation spectrum thus consists of non-interacting spinor bosons.
The ground state is determined by minimizing $E_{MF}$ with respect to
the $\bar{Q}_{ij}$ subject to the constraints
\begin{equation}
\frac{\partial E_{MF}}{\partial \bar{\lambda}_i } = 0 \, .
\end{equation}
The saddle-point value of the $\bar{Q}$ satisfies
\begin{equation}
\bar{Q}_{ij} = \langle \mathcal{J}_{\alpha\beta} b_{i}^{\alpha} b_{j}^{\beta} \,
.
\rangle
\end{equation}
Note that $\bar{Q}_{ij} = - \bar{Q}_{ji}$ indicating that $\bar{Q}_{ij}$ is a
directed field - an orientation has to be chosen on every link.

We now describe the ground state configurations of the $\bar{Q}$,
$\bar{\lambda}$ fields and the nature of the bosonic eigenspectrum for the $J_1$-$J_2$-$J_3$
model.
We examined the values of the energy
$E_{MF}$ for $\bar{Q}_{ij}$ configurations which had a translational
symmetry with two sites per unit cell. For all parameter values
configurations with a single site per unit cell were always found to
be the global minima. We will therefore restrict our attention to
such configurations. The $\bar{\lambda}_i$ field is therefore independent of
$i$, while there are six independent values of $\bar{Q}_{ij}$:
\begin{eqnarray}
\bar{Q}_{i,i+\hat{x}} &\equiv& Q_{1,x} \nonumber \\
\bar{Q}_{i,i+\hat{y}} &\equiv& Q_{1,y} \nonumber \\
\bar{Q}_{i,i+\hat{y}+\hat{x}} &\equiv& Q_{2,y+x} \nonumber \\
\bar{Q}_{i,i+\hat{y}-\hat{x}} &\equiv& Q_{2,y-x} \nonumber \\
\bar{Q}_{i,i+2\hat{x}} &\equiv& Q_{3,x} \nonumber \\
\bar{Q}_{i,i+2\hat{y}} &\equiv& Q_{3,y} \, .
\end{eqnarray}
For this choice, the bosonic eigenstates are also eigenstates of
momentum with momenta $\vec{k}$ extending over the entire first Brillouin
zone. The bosonic eigenenergies are given by
\begin{eqnarray}
\omega_{\vec{k}} &=& \left( \bar{\lambda}^2 - |A_{\vec{k}}|^2 \right)^{1/2}
\nonumber\\
A_{\vec{k}} &=& J_1 \left( Q_{1,x}\sin k_x + Q_{1,y}\sin k_y
\right)
\nonumber\\
&&+J_2 \left(
Q_{2,y+x}\sin (k_y + k_x ) +Q_{2,y-x}\sin (k_y - k_x )
\right)
\nonumber\\
&&+J_3 \left(
 Q_{3,x}\sin (2k_x ) + Q_{3,y}\sin (2 k_y ) \right) \, .
\label{omegak}
\end{eqnarray}

\begin{figure}
\centerline{\includegraphics[width=4.8in]{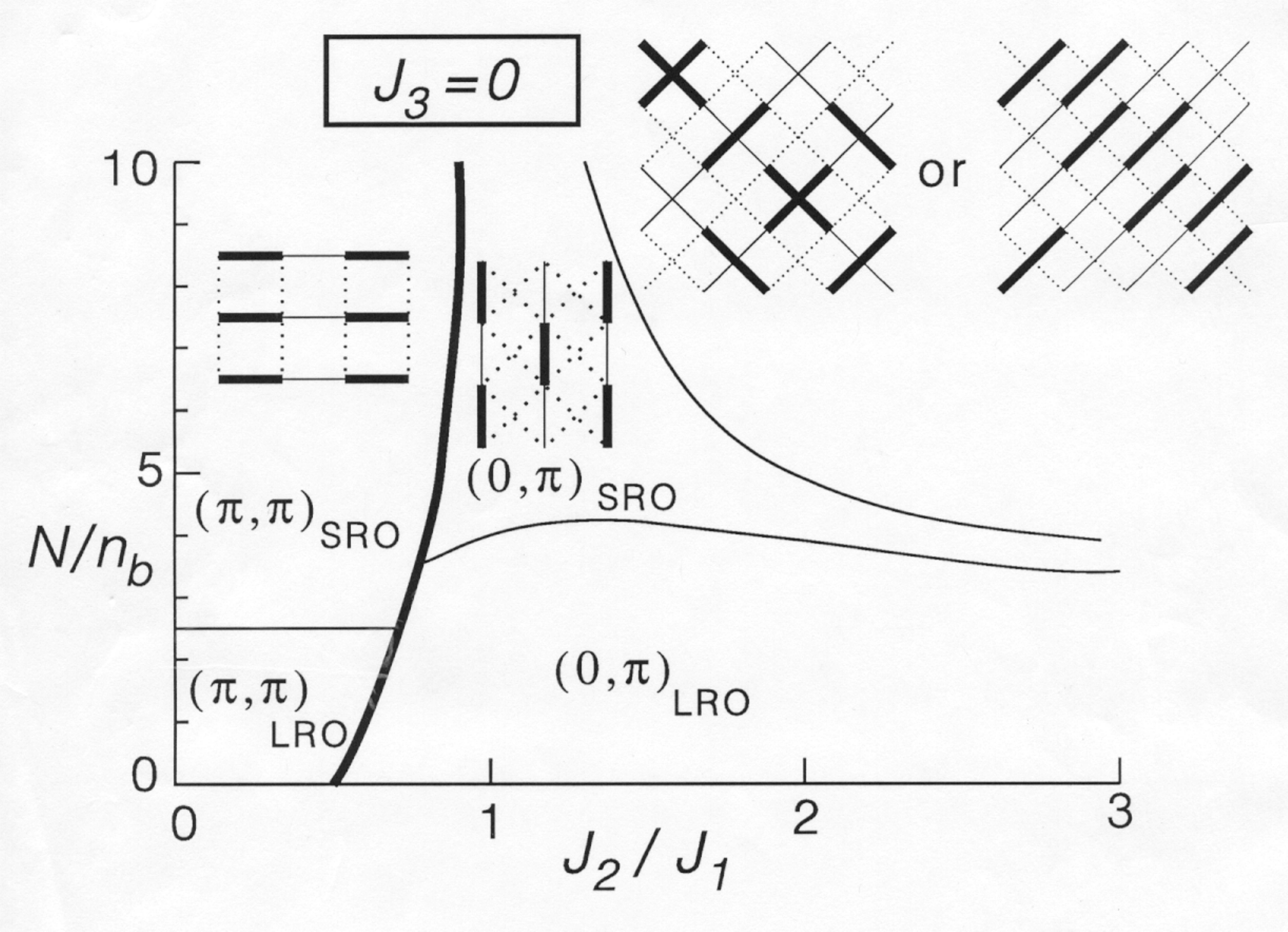}}
\caption{
Ground states of the $J_1 - J_2 - J_3$ model for $J_3=0$
as a function of $J_2/J_1$
and $N/n_b$ ($n_b = 2S$ for $SU(2)$). Thick
(thin) lines denote first (second) order transitions at $N=\infty$.
Phases are identified by the wavevectors at which they have
magnetic long-range-order (LRO) or short-range-order (SRO); the SRO
phases are ``quantum disordered'' as in Fig.~\ref{phasediag}.
The links with $Q_p\neq 0$ in each SRO phase are shown. The
large $N/n_b$, large $J_2 / J_1$ phase has the two sublattices
decoupled at $N=\infty$.
All LRO phases above have two-sublattice collinear N\'eel order.
All the SRO phases above have
valence bond solid (VBS) order at finite $N$ for odd $n_b$; this is
illustrated by the thick, thin and dotted lines.}
\label{sp2f1}
\end{figure}

\begin{figure}
\centerline{\includegraphics[width=4.8in]{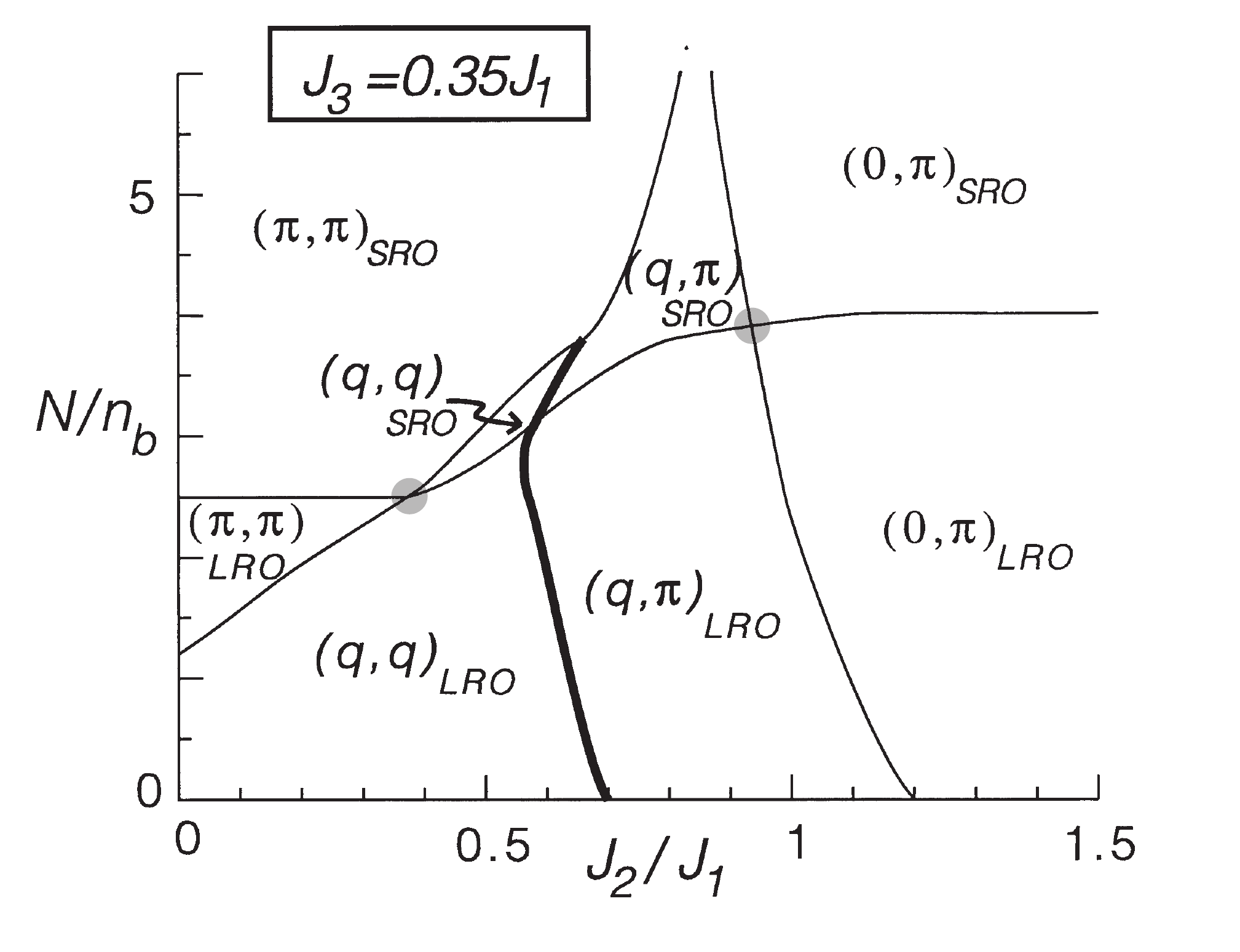}}
\caption{
As in Fig.~\ref{sp2f1}, but for $J_3 / J_1 = 0.35$. The $(0,\pi)_{SRO}$ and $(\pi,\pi)_{SRO}$ phases
have VBS order as illustrated in Fig.~\ref{sp2f1}. The $(q,q)_{SRO}$ and $(q,\pi)_{SRO}$ phases
are $Z_2$ spin liquids: they have topological order, and a topological 4-fold degeneracy of the ground state
on the torus. The $Z_2$ spin liquids here also have Ising-nematic order, {\em i.e.\/}, they break the 90$^\circ$
rotation symmetry of the square lattice, which leads to an additional 2-fold degeneracy.
The $(q,q)_{LRO}$ and $(q,\pi)_{LRO}$ have magnetic long-range order in the form of
an incommensurate spiral. The 2 shaded circles indicate regions which map onto the generalized
phase diagram in Fig.~\ref{fig:cenke}.
}
\label{sp2f3}
\end{figure}

We have numerically examined the global minima of $E_{MF}$ as a
function of the three parameters $J_2 /J_1$, $J_3 / J_1$, and
$N/n_b$~\cite{self4,self5}.
The values of the $\bar{Q}_{ij}$ at any point in the phase
diagram can then be used to classify the distinct classes of states.
The results are summarized in Figs.~\ref{sp2f1} and \ref{sp2f3} which show two
sections of the three-dimensional phase diagram. All of the phases
are labeled by the wavevector at which the spin
structure factor has a maximum. This maximum is a delta function for
the phases with magnetic LRO, while it is simply a smooth function of
$\vec{k}$ for the
quantum disordered phases (denoted by SRO in Figs.~\ref{sp2f1} and \ref{sp2f3}). The
location of this maximum will simply be twice the wavevector at which
$\omega_{\vec{k}}$ has a mimimum: this is because the structure factor
involves the product of two bosonic correlation functions, each of
which consists of a propagator with energy denominator
$\omega_{\vec{k}}$.

Each of the phases described below has
magnetic LRO for large $n_b / N$ and is quantum disordered for small
$n_b /N$. The mean-field result for the structure of all of the quantum
disordered phases is also quite simple: they are featureless spin
fluids with free spin-1/2 bosonic excitations (``spinons'') with energy
dispersion $\omega_{\vec{k}}$ which is gapped over the entire Brillouin
zone. Some of the quantum disordered phases break the lattice rotation symmetry
(leading to `Ising-nematic' order)
even at $N=\infty$ -- these will be
described below.
The mininum energy spinons lie at a wavevector $\vec{k}_0$ and
$\omega_{\vec{k}_0}$ decreases as $n_b / N$. The onset of magnetic LRO
occurs at the value of $n_b /N$ at which the gap first vanishes:
$\omega_{\vec{k}_0} = 0$. At still larger values of $n_b / N$, we get
macroscopic bose condensation of the $b$ quanta at the wavevector
$\vec{k}_0$, leading to magnetic LRO at the wavevector $2 \vec{k}_0$.

We now turn to a description of the various phases obtained. They can
be broadly classified into two types:

\subsubsection*{\underline{\sl Commensurate collinear phases}}
In these states the wavevector $\vec{k}_0$ remains pinned at a
commensurate point in the Brillouin zone, which is independent of the
values of $J_2 / J_1$, $J_3 / J_1$ and $n_b /N$. In the LRO phase, the
spin condensates on the sites are either parallel or anti-parallel to
each other, which we identify as collinear ordering. This implies
that the LRO phase remains invariant under rotations about the
condensate axis and the rotation symmetry is not completely broken.

Three distinct realizations of such states were found
\subsubsection*{a. $ (\pi , \pi )$}
This is the usual two-sublattice N\'eel state of the unfrustrated
square lattice and its quantum-disordered partner.
These states have
\begin{equation}
Q_{1,x}=Q_{1,y}\neq 0,~
Q_{2,y+x}=Q_{2,y-x}=Q_{3,x}=Q_{3,y}=0 \, .
\label{pipi}
\end{equation}
From (\ref{omegak}), the
minimum spinon excitation occurs at $\vec{k}_0 = \pm (\pi /2 , \pi /2)$.
The SRO states have no broken symmetry at $N=\infty$.
The boundary between the LRO and SRO phases occurs at
$N/n_b < 2.5$,
independent of $J_2 / J_1$ (Fig.~\ref{sp2f1}). This last feature is surely
an artifact of the large $N$ limit.
Finite $N$ fluctuations should be stronger as
$J_2 / J_1$ increases, causing the boundary to bend a little
downwards to the
right.

\subsubsection*{b. $(\pi , 0)
$ or $ (0, \pi )$}
The $(0,\pi)$ states have
\begin{equation}
Q_{1,x}=0, Q_{1,y} \neq 0,
Q_{2,y+x} = Q_{2,y-x} \neq 0,\mbox{~and~} Q_{3,x} = Q_{3,y } =0
\end{equation}
and minimum energy spinons at $\vec{k}_0 = \pm ( 0, \pi /2)$.
The degenerate
$(\pi , 0)$ state is obtained with the mapping $x
\leftrightarrow y$. The SRO state has a two-field degeneracy due to
the broken $x \leftrightarrow y$ lattice symmetry: the order associated with this symmetry
is referred to as `Ising-nematic' order. We can use the $Q$ variables here to define
an Ising nematic order parameter
\begin{equation}
\mathcal{I} = |Q_{1x}|^2 - |Q_{1y}|^2 . \label{isingnematic1}
\end{equation}
This is a gauge-invariant quantity, and the square lattice symmetry of the Hamiltonian
implies that $\langle \mathcal{I} \rangle = 0$ unless the symmetry is spontaneously broken.
The sign of $\langle \mathcal{I} \rangle$ chooses between the $(\pi, 0)$ and $(0,\pi)$ states.
The LRO state again has two-sublattice collinear N\'eel order, but the
assignment of the sublattices is different from the $(\pi , \pi )$
state. The spins are parallel along the $x$-axis, but anti-parallel
along the $y$-axis.

An interesting feature of the LRO state here is the occurrence of
``order-from-disorder''~\cite{henley}. The classical limit
($n_b / N = \infty$) of
this model has an accidental degeneracy for
$J_2 / J_1 > 1/2$: the ground state
has independent collinear N\'{e}el order on each of the $A$ and $B$
sublattices, with the energy independent of the angle between the
spins on the two sublattices. Quantum fluctuations are included
self-consistently in the $N=\infty$, $n_b / N$ finite, mean-field
theory, and lead to an alignment of the spins on the sublattices and
LRO at $(0 , \pi)$. The orientation of the ground state has thus been
selected by the quantum fluctuations.

The $(0 , \pi)$ states are
separated from the $(\pi , \pi)$ states by a first-order transition.
In particular, the spin stiffnesses of both states remain finite at
the boundary between them. This should be distinguished from the
classical limit in which the stiffness of both states vanish at their
boundary $J_2 = J_1 /2$; the finite spin stiffnesses are thus
another manifestation of order-from-disorder. At a point well away
from the singular point $J_2 = J_1 /2$, $n_b / N = \infty$ in Fig.~\ref{sp2f1},
the stiffness of both states is of order $N ( n_b / N)^2$ for
$N=\infty$ and large $n_b / N$; near this singular point however the
stiffness is of order $N ( n_b / N)$ and is induced purely by quantum
fluctuations. These results have also been obtained by a
careful resummation of the semiclassical expansion~\cite{chubukovk,mila}.

\subsubsection*{c. ``Decoupled''}
For $J_2 / J_1 $  and $N/n_b$ both
large, we have a ``decoupled'' state (Fig.~\ref{sp2f1})
with
\begin{equation}
Q_{2,y+x} = Q_{2,y-x} \neq 0~\mbox{and}~
Q_1=Q_3=0.
\end{equation}
In this case
$Q_p$ is non-zero only between sites on the same
sublattice. The two sublattices have N\'eel type SRO
which will be coupled by finite $N$ fluctuations.
The $N=\infty$
state does not break any lattice symmetry.
This state has no LRO partner.

\subsubsection*{\underline{\sl Incommensurate phases}}
In these phases the wavevector $\vec{k}_0$ and the location of the
maximum in the structure factor move continuously with the
parameters.
The spin-condensate rotates with a period which is not
commensurate with the underlying lattice spacing. Further the spin
condensate is {\em coplanar\/}: the spins rotate within a given plane
in spin space and are not collinear. There is no spin rotation axis
about which the LRO state remains invariant.

Further, no states in which the spin condensate was fully three
dimensional (``double-spiral'' or chiral states) were found; these
would be associated with complex values of $Q_p$. All the saddle
points possesed a gauge in which all the $Q_p$ were real.
Time-reversal symmetry was therefore always preserved in all the SRO
phases of Figs.~\ref{sp2f1} and \ref{sp2f3}.

The incommensurate phases occur only in models with a finite $J_3$
(Fig.~\ref{sp2f3}), at least at $N=\infty$.
There were two realizations:
\subsubsection*{d. $(\pi , q)$ or $(q, \pi)$}
Here $q$ denotes a wavevector which varies continuously between $0$
and $\pi$ as the parameters are changed. The $(q , \pi )$ state has
\begin{equation}
Q_{1,x}\neq Q_{1,y} \neq 0,~
Q_{2,x+y} = Q_{2,y-x} \neq 0,~Q_{3,x} \neq 0~\mbox{and}~
Q_{3,y}=0;
\end{equation}
the
degenerate $(\pi,q)$ helix is obtained by the mapping $x
\leftrightarrow
y$. The SRO state has a two-fold degeneracy due to the broken $x
\leftrightarrow y$ lattice symmetry, and so this state has Ising-nematic order.
The order parameter in Eq.~(\ref{isingnematic1}) continues to measure this broken symmetry.

\subsubsection*{e. $(q, q)$ or $(q, -q)$}
The $(q,q)$ state has
\begin{equation}
Q_{1,x}=Q_{1,y} \neq 0,~
Q_{2,x+y}\neq 0,~Q_{2,y-x} = 0,~Q_{3,x}=Q_{3,y}\neq 0;
\end{equation}
this
is degenerate with the $(q, -q)$ phase. The SRO state therefore has a
two-fold degeneracy due to a broken lattice reflection symmetry, and so it also has
Ising nematic order. However, the Ising symmetry now corresponds to reflections about
the principle square axes, and the analog of Eq.~(\ref{isingnematic1}) is now
\begin{equation}
\mathcal{I} = |Q_{2,x+y}|^2 - |Q_{2,y-x}|^2 .
\label{isingnematic2}
\end{equation}

As we noted above, the broken discrete symmetries in states
with SRO at $(0 , \pi )$ and $(q , \pi)$ are identical:
both are two-fold degenerate due to a breaking of the
$x \leftrightarrow y$ symmetry. The states
are only distinguished by a non-zero value of $Q_3$
in
the $(q, \pi )$ phase and the accompanying incommensurate
correlations in the spin-spin correlation functions.
However $Q_3$ is gauge-dependent
and so somewhat
unphysical as an order parameter.
In the absence of any further fluctuation-driven lattice
symmetry
breaking, the transition between SRO at $(0, \pi
)$
and $(q , \pi)$ is an example of a {\em disorder
line}~\cite{disorder}; these are lines at which
incommensurate correlations first turn on. However, we will see that
quantum fluctuations clearly distinguish these two phases, which have
confined and deconfined spinons respectively, and the associated topological
order requires a phase transitions between them.

An interesting feature of Fig.~\ref{sp2f3} is that the
commensurate
states squeeze out the incommensurate phases as $N/n_b$
increases.
We expect that this suppression of incommensurate order by
quantum
fluctuations is
a general feature of frustrated
antiferromagnets.

\subsubsection{Fluctuations -- long wavelength effective
actions}
\label{sec:6B}

We now extend the analysis of Section~\ref{sec:6A} beyond the mean-field theory
and examine the consequences of corrections at finite $N$. The main
question we hope to address are:
\begin{itemize}
\item
The mean-field theory yielded an excitation spectrum consisting of
free spin-1/2 bo\-son\-ic spinons. We now want to understand the nature of
the forces between these spinons and whether they can lead to
confinement of half-integer spin excitations.
\item
Are there any collective excitations and does their dynamics modify in
any way the nature of the mean field ground state ?
\end{itemize}

The structure of the fluctuations will clearly be determined by the
low-energy excitations about the mean-field state. We have already
identified one set of such excitations: spinons at momenta near
mimima in their dispersion spectrum, close to the onset of the magnetic
LRO phase whence the spinon gap vanishes. An additional set of
low-lying spinless excitations can arise from the fluctuations of the
$Q_{ij}$ and $\lambda_i$ fields about their mean-field values. The
gauge-invariance (\ref{gaugetrans}) will act as a powerful restriction
on the allowed terms in the effective action for these spinless fields. We
anticipate that the only such low-lying excitations are associated
with the $\lambda_i$ and the {\em phases\/} of the $Q_{ij}$. We
therefore
parametrize
\begin{equation}
Q_{i,i+\hat{e}_p} = \bar{Q}_{i,i+\hat{e}_p}
\exp\left( -i \Theta_p \right) \, ,
\label{qtheta}
\end{equation}
where the vector $\hat{e}_p$ connects the two sites of the lattice
under consideration,
$\bar{Q}$ is the mean-field value, and $\Theta_p$ is a real phase.
The gauge invariance (\ref{gaugetrans}) implies that the
effective action for the $\Theta_p$ must be invariant under
\begin{equation}
\Theta_p \rightarrow \Theta_p + \rho_i + \rho_{i+\hat{e}_p} \, .
\end{equation}
Upon performing a Fourier transform, with the link variables $\Theta_p$
placed on the center of the links, the gauge invariance takes the
form
\begin{equation}
\Theta_p ( \vec{k} ) \rightarrow
\Theta_p (  \vec{k} ) + 2 \rho (\vec{k}  ) \cos( k_p /2) \, ,
\label{thetatr}
\end{equation}
where $k_p = \vec{k} \cdot \hat{e}_p$.
This invariance implies that the effective action
for the $\Theta_p$, after integrating out the $b$ quanta,
can only be a function of the following gauge-invariant
combinations:
\begin{equation}
I_{pq} = 2 \cos(k_q /2) \Theta_p (\vec{k}  ) -
2 \cos (k_p /2) \Theta_q (\vec{k}  ) \, .
\label{ipqdef}
\end{equation}
We now wish to take the continuum limit at points in the Brillouin zone
where the action involves only gradients of the $\Theta_p$ fields and
thus has the possibility of gapless excitations. This involves
expanding about points in the Brillouin zone where
\begin{equation}
\cos ( k_p / 2) = 0~\mbox{for the largest numbers of $\hat{e}_p$}.
\label{lowenergy}
\end{equation}
We now apply this general principle to the $J_1$-$J_2$-$J_3$ model.

\subsubsection*{\underline{\sl Commensurate collinear phases}}
We begin by examining the $(\pi , \pi)$-SRO phase.
As noted in (\ref{pipi}), this phase has the mean field
values
$Q_{1,x} = Q_{1,y} \neq 0$, and all other $\bar{Q}_{ij}$ zero.
Thus we need only examine the condition (\ref{lowenergy}) with
$\hat{e}_p = \hat{e}_x , \hat{e}_y$. This uniquely identifies the
point $\vec{k} = \vec{G} = (\pi , \pi)$ in the Brillouin zone. We therefore
parametrize
\begin{equation}
\Theta_x ( \vec{r} ) =  e^{i \vec{G} \cdot \vec{r}} A_x ( \vec{r} )
\label{thetax}
\end{equation}
and similarly for $\Theta_y$; it can be verified that both $\Theta$
and $A_x$ are real in the above equation.
We will also be examining invariances of
the theory under gauge transformations near $\vec{G}$: so we write
\begin{equation}
\rho ( \vec{r} ) = e^{i \vec{G} \cdot \vec{r} } \zeta ( \vec{r} ) \, .
\label{rhophi}
\end{equation}
It is now straightforward to verify that the gauge transformations
(\ref{thetatr}) are
equivalent to
\begin{equation}
A_x \rightarrow A_x + \partial_x \zeta
\end{equation}
and similarly for $A_{y}$. We will also need in the continuum limit
the component of $\lambda$ near the wavevector $\vec{G}$. We therefore
write
\begin{equation}
i \lambda_i = \bar{\lambda} + i e^{i \vec{G} \cdot \vec{r}}  A_{\tau} (
\vec{r}_i ) \, .
\label{anslam}
\end{equation}
Under gauge transformations we have
\begin{equation}
A_{\tau} \rightarrow A_{\tau} + \partial_{\tau} \zeta \, .
\end{equation}
Thus $A_x$, $A_y$, $A_{\tau}$ transform as components of a continuum
$U(1)$ vector gauge field.

We will also need the properties of the boson operators under
the gauge transformation $\zeta$. From (\ref{gaugetrans}) and (\ref{rhophi}) we see that the
bosons on the two sublattices ($A,B$) carry opposite charges $\pm 1$:
\begin{eqnarray}
b_{A} &\rightarrow& b_{A} e^{i\zeta} \nonumber \\
b_{B} &\rightarrow& b_{B} e^{-i\zeta} \, .
\end{eqnarray}
Finally, we note that the bosonic eigenspectrum has a minimum near $\vec{k}
= \vec{k}_0 = (\pi/2 , \pi/2 )$; we therefore parametrize
\begin{eqnarray}
b_{Ai}^{\alpha} &=& \psi_1^{\alpha} (\vec{r}_i ) e^{i \vec{k}_0 \cdot \vec{r}_i }
\nonumber \\
b_{Bi}^{\alpha} &=& -i \mathcal{J}^{\alpha\beta} \psi_{2\beta} (\vec{r}_i )
e^{i \vec{k}_0 \cdot \vec{r}_i } \, .
\label{bosepar}
\end{eqnarray}

We insert the continuum parameterizations (\ref{thetax}),
(\ref{anslam})
and (\ref{bosepar}) into the functional integral (\ref{zfunct}),
perform a gradient expansion, and transform the Lagrangian ${\cal L}$ into
\begin{eqnarray}
{\cal L} &=& \int \frac{d^2 r}{a^2} \left [
\psi_{1\alpha}^{\ast} \left( \frac{d}{d\tau} + i A_{\tau}
\right)
\psi_1^{\alpha} +
\psi_{2}^{\alpha\ast} \left( \frac{d}{d\tau} - i A_{\tau}
\right)
\psi_{2\alpha} \right.
\nonumber \\
&& \qquad\qquad
 + \bar{\lambda} \left( |\psi_1^{\alpha} |^2
+ |\psi_{2\alpha} |^2 \right) 
-4 J_1 \bar{Q}_1 \left ( \psi_1^{\alpha}\psi_{2\alpha} +
\psi_{1\alpha}^{\ast}\psi_2^{\alpha\ast}
\right )
\nonumber \\
&&\qquad\qquad
+  J_1 \bar{Q}_1 a^2 \left [
\left ( \vec{\nabla}  + i  \vec{A} \right ) \psi_1^{\alpha}
\left ( \vec{\nabla}  - i \vec{A} \right ) \psi_{2\alpha} \right.
\nonumber \\
&& \qquad\qquad\qquad
+ \left.
\left ( \vec{\nabla} - i \vec{A} \right )
\psi_{1\alpha}^{\ast}
\left ( \vec{\nabla} + i \vec{A} \right )
\psi_2^{\alpha\ast} \right ]  \Biggr] \, .
\label{charge}
\end{eqnarray}
We now introduce the fields
\begin{eqnarray*}
z^{\alpha} & = & (\psi_1^{\alpha} +
\psi_2^{\alpha\ast})/\sqrt{2} \\
\pi^{\alpha} & = & (\psi_1^{\alpha} -
\psi_2^{\alpha\ast})/\sqrt{2} \, .
\end{eqnarray*}
Following the definitions of the underlying spin operators, it is not difficult to show
that the N\'eel order parameter $\varphi_a$ (which is proportional to $\vec{n}$ in (\ref{neel}))
is related to the $z_\alpha$ by
\begin{equation}
\varphi_a = z_\alpha^{\ast} {\sigma^{a\alpha}}_{\beta} z^\beta \, . \label{neelz}
\end{equation}
From Eq.~(\ref{charge}), it is clear that the
the $\pi$ fields turn out to have mass $\bar{\lambda} + 4 J_1 \bar{Q}_1$,
while the $z$ fields
have a mass $\bar{\lambda} - 4 J_1 \bar{Q}_1$ which vanishes at the
transition to the LRO
phase. The $\pi$ fields can therefore
be safely integrated out,
and ${\cal L}$ yields
the following effective action, valid at distances much
larger than the lattice
spacing~\cite{self2,self3}:
\begin{equation}
S_{\rm eff} =
\int \frac{d^2 r}{\sqrt{8}a} \int_{0}^{c\beta}
d\tilde{\tau} \left\{
|(\partial_{\mu} - iA_{\mu})z^{\alpha}|^2
+ \frac{\Delta^2}{c^2}
|z^{\alpha} |^2\right\} \,.
\label{sefp}
\end{equation}
Here $\mu$ extends over $x,y,\tau$,
$c = \sqrt{8}J_1 \bar{Q}_1 a$ is the spin-wave velocity,
$\Delta = (\bar{\lambda}^2 - 16J_1^2 \bar{Q}_1^2 )^{1/2}$ is the gap
towards spinon excitations,
and $A_{\tilde{\tau}} = A_{\tau}/c$.
Thus, in its final form, the long-wavelength theory consists
of a massive, spin-1/2, relativistic, boson $z^{\alpha}$ (spinon)
coupled to a {\em compact\/} $U(1)$ gauge field. By `compact' we mean
that values $A_\mu$ and $A_\mu + 2 \pi$ are identified with each other,
and the gauge field lives on a circle: this is clearly required by Eq.~(\ref{qtheta}).

At distances larger than $c/\Delta$, we may safely integrate out the
massive $z$ quanta and obtain a a compact $U(1)$ gauge theory in 2+1
dimensions. This theory was argued by
Polyakov~\cite{polyakovbook,polyakov} to be permanently in a
confining phase, with the confinement driven by ``monopole''
tunnelling events. The compact $U(1)$ gauge force will therefore
confine the $z^{\alpha}$ quanta in pairs.
So the conclusion is that the $(\pi,\pi)_{SRO}$ does {\em not\/}
possess $S=1/2$ spinon excitations, as was the case in the mean field
theory. Instead, the lowest-lying excitations with non-zero spin will
be triplons, similar to those in Section~\ref{sec:dimer}.
A further important effect here, not present in the $U(1)$ gauge theories
considered by Polyakov, is that the monopole tunnelling events carry Berry phases.
The influence of these Berry phases has been described \cite{self2,self3} and reviewed \cite{qpmott}
elsewhere, and so will not be explained here. The result is that the condensation
of monopoles with Berry phases leads to valence bond solid (VBS) order
in the ground state. This order is associated with the breaking of the square lattice
space group symmetry, as illustrated in Figs.~\ref{sp2f1} and~\ref{fig:cenke} below.
For the $(\pi,\pi)_{SRO}$ phase, this means that the singlet spin correlations
have a structure similar to that in Fig.~\ref{fig2}. In other words, the square lattice
antiferromagnet {\em spontaneously} acquires a ground state with a symmetry similar
to that of the paramagnetic phase of coupled-dimer antiferromagnet.
Because the VBS order is spontaneous, the ground state is 4-fold degenerate (associated
with 90$^\circ$ rotations about a lattice site), unlike the non-degenerate ground state
of the dimerized antiferromagnet of Section~\ref{sec:dimer}.
VBS states with a plaquette ordering pattern can also appear, but are not shown in the figures.

The quantum phase transition between the $(\pi,\pi)_{SRO}$ and $(\pi, \pi)_{LRO}$
phases has been the topic of extensive study. The proposal of Refs.~\cite{senthil1,senthil2}
is that monopoles are suppressed precisely at the quantum critical point, and so
the continuum action in Eq.~(\ref{sefp}) constitutes a complete description of the
critical degrees of freedom. It has to be supplemented by a quartic non-linearity
$\left( |z^\alpha |^2 \right)^2$, because such short-range interactions are relevant
perturbations at the critical point. A review of this deconfined criticality proposal
is found elsewhere \cite{solvay}.

The properties of the $(0, \pi)$ phase are very similar to those of
the $(\pi , \pi)$ phase considered above. It can be shown quite generally that any quantum
disordered state which has appreciable commensurate, collinear spin
correlations will have similar properties: confined spinons, a
collective mode described by a compact $U(1)$ gauge field, and VBS
order for odd $n_b$.

\subsubsection*{\underline{\sl Incommensurate phases}}

We now turn to a study of the incommensurate phases. It is not
difficult to show that in this case it is not possible to satisfy the
constraints (\ref{lowenergy}) at any point in the Brillouin zone for
all the non-zero $Q_p$. This implies that, unlike the commensurate phases,
there is no gapless
collective gauge mode in the gaussian fluctuations of the incommensurate SRO phases.
This has the important implication that the mean-field theory is stable:
the structure of the mean-field ground state, and its spinon excitations
will survive fluctuation corrections. Thus we obtain a stable `spin liquid' with
bosonic $S=1/2$ spinon excitations. We will now show that these spinons carry
a $Z_2$ gauge charge, and so this phase is referred to as a $Z_2$ spin liquid.
The $Z_2$ gauge field also accounts for `topological order' and a 4-fold ground
state degeneracy on the torus.

The structure of the theory is simplest in the vicinity of a
transition to a commensurate collinear phase:
we now examine the effective action as one moves
from the $(\pi , \pi)$-SRO phase into the $(q,q)$-SRO phase
(Fig.~\ref{sp2f3};
a very similar analysis can be performed at the
boundary between the $(\pi , \pi)$-SRO and the $(\pi , q)$-SRO phases).
This transition is characterized by a continuous
turning on of non-zero values of $Q_{i,i+\hat{y}+\hat{x}}$,
$Q_{i,i+2\hat{x}}$ and $Q_{i,i+2\hat{y}}$. It is easy to see from
Eq.~(\ref{gaugetrans}) that these fields transform as scalars of
charge $\pm 2$ under the gauge transformation associated
with $A_{\mu}$. Performing a gradient expansion upon the
bosonic fields coupled to these scalars we find that the
Lagrangian ${\cal L}$ of the $(\pi , \pi)$-SRO phase gets
modified to
\begin{equation}
{\cal L} \rightarrow {\cal L} + \int \frac{d^2 r}{a} \left(
\vec{\Lambda}_A \cdot \left(
\mathcal{J}_{\alpha\beta}\psi_1^{\alpha}\vec{\nabla}\psi_1^{\beta}
\right) +
\vec{\Lambda}_B \cdot \left(
\mathcal{J}^{\alpha\beta}\psi_{2\alpha}\vec{\nabla}\psi_{2\beta}
\right) + \mbox{c.c.} \right) \, ,
\end{equation}
where $\vec{\Lambda}_{A,B}$ are two-component scalars
$\equiv (J_3 Q_{3,x} + J_2 Q_{2,y+x},
J_3 Q_{3,y}+J_2$ $Q_{2,y+x} )$ with the sites on the ends of
the link
variables on sublattices $A,B$. Finally, as before, we
transform to the $z,\pi$ variables, integrate out the $\pi$
fluctuations and obtain~\cite{self5}
\begin{eqnarray}
S_{\rm eff} &=&
\int \frac{d^2 r}{\sqrt{8}a} \int_{0}^{c\beta}
d\tilde{\tau} \Biggl\{
|(\partial_{\mu} - iA_{\mu})z^{\alpha}|^2
+ s_z
|z^{\alpha} |^2 + \vec{\Lambda} \cdot \left(
\mathcal{J}_{\alpha\beta}z^{\alpha}\vec{\nabla} z^{\beta}\right) +
\mbox{c.c.} \nonumber \\
&~&~~~~~~~ ~+ K_{\Lambda} |(\partial_\mu + 2 i A_\mu) \vec{\Lambda}|^2 + s_\Lambda \vec{\Lambda}^2 + \mbox{terms quartic in $z^\alpha$, $\vec{\Lambda}$}
\Biggr\} \, .
\label{hgs}
\end{eqnarray}
Here $s_z  = \Delta^2 /c^2$, $\vec{\Lambda} = (\vec{\Lambda}_A + \vec{\Lambda}_B^{\ast} )
/(2J_1 \bar{Q}_1 a)$ is a
complex scalar of
charge $-2$, and $K_\Lambda$ is a stiffness.
We have explicitly written the quadratic terms in the effective action for the $\vec{\Lambda}$: these
are generated by short wavelength
fluctuations of the $b^{\alpha}$ quanta. We have omitted quartic and higher order terms which are needed to stabilize the theory when the
`masses' $s_z$ or $s_\Lambda$ are negative, and are also important near the quantum phase transitions. This effective action is also the simplest theory that can be written
down which couples a spin-1/2, charge 1, boson $z^{\alpha}$, a
compact $U(1)$ gauge field $A_{\mu}$, and a two spatial component,
charge $-2$, spinless boson $\vec{\Lambda}$. It is the main result of
this section and summarizes essentially
all of the physics we are trying to describe.

\begin{figure}[t]
\centering
\includegraphics[width=4in]{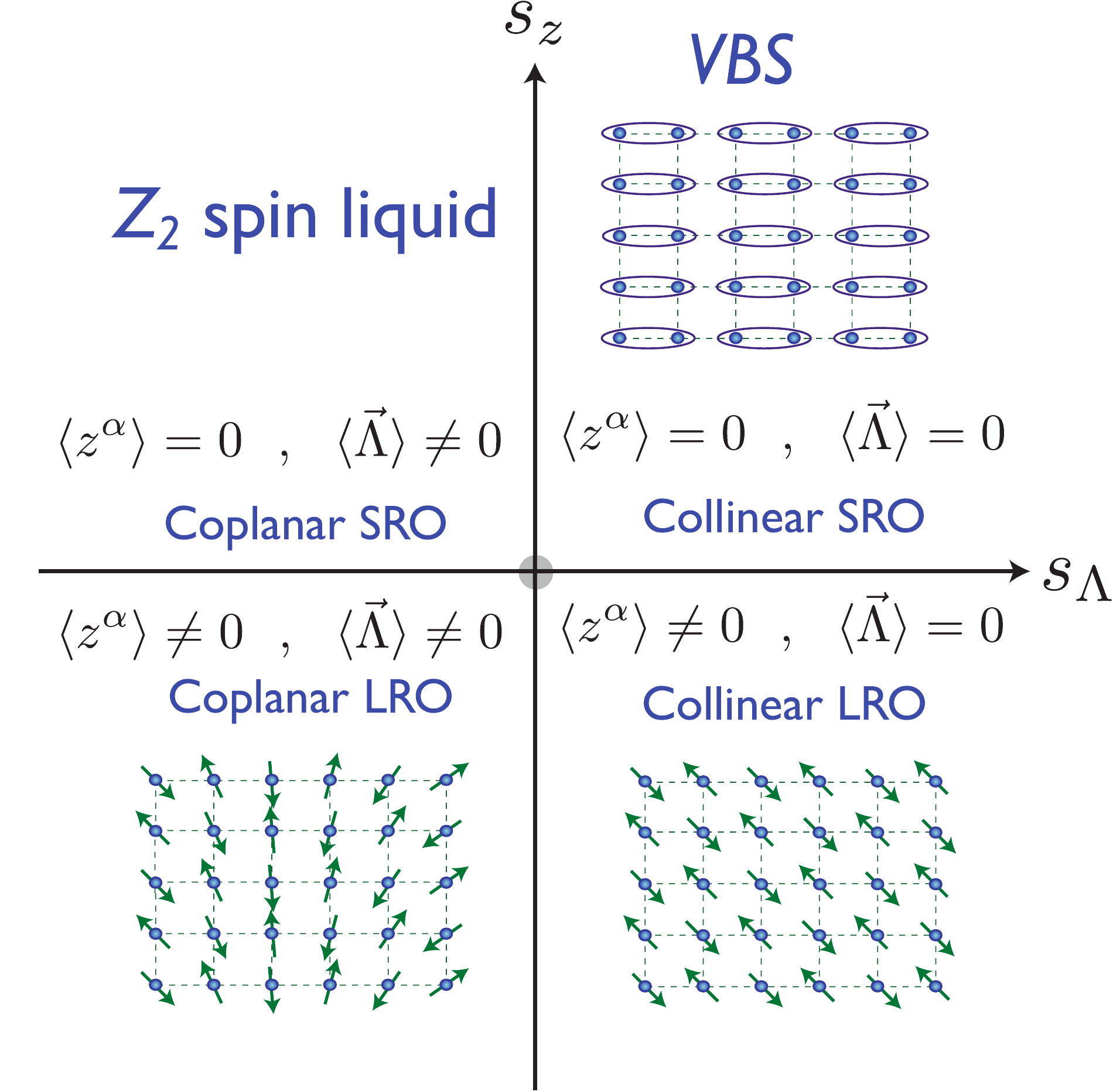}
\caption{Phase diagram of the theory $S_{\rm eff}$ in Eq.~(\ref{hgs}) for the bosonic spinons $z_\alpha$
and the charge -2 spinless boson $\vec{\Lambda}$. Fig.~\ref{sp2f1} contains examples of the region
$s_\Lambda > 0$. Fig.~\ref{sp2f3} contains 2 separate instances of 4 phases meeting at a point as above, with
the 4 phases falling into the classes labeled above; these points are labeled in both figures by the shaded circles.}
\label{fig:cenke}
\end{figure}
We now describe the various phases of $S_{\rm eff}$, which are summarized in Fig.~\ref{fig:cenke}.
\begin{enumerate}
\item
\underline{Commensurate, collinear, LRO:} $\langle z^{\alpha} \rangle
\neq 0$, $\langle \vec{\Lambda} \rangle = 0$\\
This is the $(\pi,\pi)_{LRO}$ state with commensurate, collinear, magnetic LRO.
\item
\underline{Commensurate, collinear, SRO:} $\langle z^{\alpha} \rangle
= 0$, $\langle \vec{\Lambda} \rangle = 0$\\
This is the $(\pi,\pi)_{SRO}$ ``quantum-disordered'' state with collinear spin correlations
peaked at $(\pi , \pi)$. Its properties were described at length
above. The compact $U(1)$ gauge force confines the $z^{\alpha}$
quanta. The spinless collective mode associated with the gauge
fluctuations acquires a gap from monopole condensation, and the monopole
Berry phases induce VBS order for odd $n_b$.
\item
\underline{Incommensurate, coplanar, SRO:} $\langle z^{\alpha} \rangle
= 0$, $\langle \vec{\Lambda} \rangle \neq 0$\\
This is the incommensurate phase with SRO at $(q,q)$ which we want
to study. It is easy to see that condensation of $\vec{\Lambda}$
necessarily implies the appearance of incommensurate SRO:
ignore fluctuations of $\vec{\Lambda}$ about $\langle \vec{\Lambda} \rangle$
and diagonalize the quadratic form controlling the $z^{\alpha}$
fluctuations; the minimum of the
dispersion
of the $z^{\alpha}$ quanta is at a non-zero wavevector
\begin{equation}
\vec{k}_0 =
(\langle\Lambda_x\rangle
, \langle\Lambda_y\rangle )/2 \, .
\end{equation}
The spin structure factor will therefore have a maximum at an
incommensurate wave\-vec\-tor. This phase also has a broken lattice
rotation symmetry due to the choice of orientation in the $x-y$ plane
made by $\vec{\Lambda}$ condensate, {\em i.e.\/}, it has Ising-nematic order.\\
The condensation of $\vec{\Lambda}$ also has a dramatic impact on the
nature of the force between the massive $z^{\alpha}$ quanta. Detailed
arguments have been presented by Fradkin and Shenker~\cite{fradshen} that the
condensation of a doubly charged Higgs scalar quenches the confining
compact $U(1)$ gauge force in 2+1 dimensions between singly charged
particles. We can see this here from Eq.~(\ref{hgs})
by noticing that the condensation of $\vec{\Lambda}$ expels $A_\mu$
by the Meissner effect: consequently, monopoles in $A_\mu$ are connected
by a flux tube whose action grows linearly with the separation between monopoles.
The monopoles are therefore confined, and are unable to induce
the confinement of the $z^\alpha$ quanta. From Eq.~(\ref{hgs}) we also see
that once $\vec{\Lambda}$ is condensed, the resulting theory for the spinons
only has an effective $Z_2$ gauge invariance: Ref.~\cite{fradshen} argued that
there is an effective description of this free spinon phase in terms of a $Z_2$ gauge theory.
The excitation structure is therefore very similar to that of the
mean-field theory: spin-1/2, massive bosonic spinons and spinless
collective modes which have a gap. The collective mode gap is present
in this case even at $N=\infty$ and is associated with the
condensation of $\vec{\Lambda}$.\\
This state is also `{\em topologically\/} ordered'. We can see this by noticing \cite{self4,wen1} that it carries
stable point-like excitations which are $ 2 \pi$ vortices in either of $\Lambda_x$ or $\Lambda_y$.
Because of the screening by the $A_\mu$ gauge field, the vortices carry a finite energy (this is analogous
to the screening of supercurrents by the magnetic field around an Abrikosov vortex in a superconductor).
Because the $\vec{\Lambda}$ carry charge $-2$, the total $A_\mu$ flux trapped by a vortex is $\pi$.
Thus the vortices are also stable to monopole tunneling events, which change the $A_\mu$ flux by integer
multiples of $2 \pi$. A $z_\alpha$ spinon circumnavigating such a vortex would pick up an Aharanov-Bohm
phase factor of $\pi$ (because the spinons have unit charge), and this is equivalent to the
statement that the vortex and the spinon obey mutual `semionic' statistics. All these characteristics identify
the vortex excitation as one dubbed later \cite{sf} as the {\em vison\/}.\\
The vison also allows us to see the degeneracy of the gapped ground state on surfaces of non-trivial topology.
We can insert a vison through any of the `holes' in surface, and obtain a new candidate eigenstate.
This eigenstate has an energy essentially degenerate with the ground state because the core of the vortex
is within the hole, and so costs no energy. The `far field' of the vison is within the system, but it costs negligible energy
because the currents have been fully screened by $A_\mu$ in this region. Thus we obtain
a factor of 2 increase in the degeneracy for every `hole' in the surface (which is in turn related to the genus
of the surface).\\
The state so obtained is now referred to as a $Z_2$ spin liquid, and has been labeled as such in the figures.
As we noted above, the present theory only yields $Z_2$ spin liquids with Ising-nematic order,
associated with broken symmetry of 90$^\circ$ lattice rotations.
We also note an elegant exactly solvable model described by Kitaev \cite{kitaev}, which has spinon and vison excitations
with the characteristics described above, but without the Ising-nematic order.
\item
\underline{Incommensurate, coplanar, LRO:} $\langle z^{\alpha} \rangle
\neq 0$, $\langle \vec{\Lambda} \rangle \neq 0$\\
The condensation of the $z$ quanta at the wavevector $\vec{k}_0$ above
leads to incommensurate LRO in the $(q,q)_{LRO}$ phase, with the spin condensate spiraling in
the plane.
\end{enumerate}
We also note a recent work \cite{cenke,solvay,qiprl} which has given a dual perspective on the above
phases, including an efficient description of the phase transitions between them, and applied the results
to experiments on $\kappa$-(ET)$_2$Cu$_2$(CN)$_3$.

\section{$d$-wave superconductors}
\label{sec:dwave}

In our discussion of phase transitions in insulators we found
that the low-energy excitations near the critical point were
linked in some way to the broken symmetry of the magnetically
ordered state. In the models of Section~\ref{sec:dimer} the low-energy
excitations involved long wavelength fluctuations of the
order parameter. In Section~\ref{sec:frust} the connection to the
order parameter was more subtle but nevertheless present: the field $z_\alpha$
in Eq.~(\ref{hgs}) is a `fraction' of the order parameter as indicated in (\ref{neelz}),
and the gauge field $A_\mu$ represents a non-coplanarity in the local
order parameter orientation.

We will now move from insulators to the corresponding transitions in $d$-wave
superconductors. Thus we will directly address the criticality of the magnetic QPT
at $x=x_s$ in Fig.~\ref{fig:figglobal}. We will also consider the
criticality of the `remnant' Ising-nematic ordering at $x=x_m$ within the
superconducting phase. A crucial property of $d$-wave superconductors is that they
generically contain gapless, fermionic Bogoliubov excitations, as we will review below.
These gapless excitations have a massless Dirac spectrum near isolated points in Brillouin zone.
While these fermionic excitations are present in the non-critical $d$-wave superconductor, it is natural
to ask whether they modify the theory of the QPT. Even though they may not be
directly related to the order parameter, we can ask if the order parameter and fermionic excitations
couple in interesting ways, and whether this coupling modifies the universality class of the transition.
These questions will be answered in the following subsections.

We note that symmetry breaking transitions in graphene are also described by field theories
similar to those discussed in this section \cite{ssherbut,ssherbut2}.

\subsection{Dirac fermions}
\label{sec:dirac}
We begin with a review of the standard BCS mean-field
theory for a $d$-wave superconductor on the square lattice, with an eye towards identifying
the fermionic Bogoliubov quasiparticle excitations. For now, we assume we are far from
any QPT associated with SDW, Ising-nematic, or other broken symmetries.
We
consider the generalized Hamiltonian
\begin{equation}
H_{tJ} = \sum_{k} \varepsilon_k c_{k \alpha}^{\dagger} c_{k
\alpha} + J_1 \sum_{\langle ij \rangle} \vec{S}_i \cdot \vec{S}_{j} \, ,
\label{g2}
\end{equation}
where $c_{j\alpha}$ is the annihilation operator for an electron
on site $j$ with spin $\alpha=\uparrow,\downarrow$, $c_{k\alpha}$
is its Fourier transform to momentum space, $\varepsilon_k$ is the
dispersion of the electrons (it is conventional to choose
$\varepsilon_k = -2t_1 (\cos(k_x) + \cos(k_y)) - 2 t_2 ( \cos(k_x
+ k_y) + \cos(k_x - k_y)) - \mu$, with $t_{1,2}$ the first/second
neighbor hopping and $\mu$ the chemical potential), and the $J_1$
term is the same as that in Eq.~(\ref{hamil}) with
\begin{equation}
S_{ja} = \frac{1}{2} c^{\dagger}_{j\alpha}
\sigma_{\alpha\beta}^{a} c_{j \beta} \label{g2a}
\end{equation}
and $\sigma^{a}$ the Pauli matrices. We will consider
the consequences of the further neighbor exchange interactions in (\ref{hamil}) for the
superconductor in Section~\ref{sec:did} below. Applying the BCS
mean-field decoupling to $H_{tJ}$ we obtain the Bogoliubov
Hamiltonian
\begin{equation}
H_{BCS} = \sum_{k} \varepsilon_k c_{k \alpha}^{\dagger} c_{k
\alpha} - \frac{J_1}{2} \sum_{j\mu}\Delta_{\mu} \left(
c^{\dagger}_{j\uparrow} c^{\dagger}_{j+\hat{\mu},\downarrow} -
c^{\dagger}_{j\downarrow} c^{\dagger}_{j+\hat{\mu},\uparrow}
\right) + \mbox{h.c.} \, . \label{g3}
\end{equation}
For a wide range of parameters, the ground state energy is optimized
by a $d_{x^2-y^2}$ wavefunction for the Cooper pairs: this
corresponds to the choice $\Delta_x = - \Delta_y =
\Delta_{x^2-y^2}$. The value of $\Delta_{x^2-y^2}$ is determined
by minimizing the energy of the BCS state
\begin{equation}
E_{BCS} = J_1 |\Delta_{x^2-y^2}|^2 - \int \frac{d^2 k}{4 \pi^2}
\left[ E_k - \varepsilon_k \right] \, , \label{g4}
\end{equation}
where the fermionic quasiparticle dispersion is
\begin{equation}
E_k = \left[ \varepsilon_k^2 + \left|J_1 \Delta_{x^2-y^2}(\cos k_x
- \cos k_y)\right|^2 \right]^{1/2}. \label{g5}
\end{equation}

The energy of the quasiparticles, $E_k$, vanishes at the four
points $(\pm Q, \pm Q)$ at which $\varepsilon_k=0$. We are
especially interested in the low-energy quasiparticles in the
vicinity of these points, and so we perform a gradient expansion
of $H_{BCS}$ near each of them. We label the points $\vec{Q}_1=(Q,Q)$,
$\vec{Q}_2=(-Q,Q)$, $\vec{Q}_3=(-Q,-Q)$, $\vec{Q}_4=(Q,-Q)$ and write
\begin{equation}
c_{j\alpha} =f_{1\alpha} (\vec{r}_j) e^{i \vec{Q}_1 \cdot \vec{r}_j}+f_{2\alpha} (\vec{r}_j)
e^{i \vec{Q}_2 \cdot \vec{r}_j}+f_{3\alpha} (\vec{r}_j) e^{i \vec{Q}_3 \cdot \vec{r}_j}+f_{4\alpha} (\vec{r}_j)
e^{i \vec{Q}_4 \cdot \vec{r}_j},\label{g5a}
\end{equation}
while assuming that the $f_{1-4,\alpha} (\vec{r})$ are slowly varying
functions of $\vec{r}$. We also introduce the bispinors $\Psi_1 =
(f_{1\uparrow}, f_{3\downarrow}^{\dagger},
f_{1\downarrow},-f_{3\uparrow}^{\dagger})$, and $\Psi_2 =
(f_{2\uparrow}, f_{4\downarrow}^{\dagger},
f_{2\downarrow},-f_{4\uparrow}^{\dagger})$, and then express
$H_{BCS}$ in terms of $\Psi_{1,2}$ while performing a spatial
gradient expansion. This yields the following effective action for
the fermionic quasiparticles:
\begin{eqnarray}
&& \mathcal{S}_{\Psi} = \int d \tau d^2 r \Biggl[
\Psi_{1}^{\dagger}  \left(
\partial_\tau -i \frac{v_F}{\sqrt{2}} (\partial_x + \partial_y) \tau^z  -i \frac{v_\Delta}{\sqrt{2}} (-\partial_x +
\partial_y) \tau^x \right) \Psi_{1}   \nonumber \\
&&~~~~~~~~~~~~~~~~~~+  \Psi^\dagger_2 \left(
\partial_\tau - i \frac{v_F}{\sqrt{2}} (-\partial_x + \partial_y) \tau^z -i  \frac{v_\Delta}{\sqrt{2}} (\partial_x + \partial_y)  \tau^x \right) \Psi_{2} \Biggr]
\, , \label{dsid1}
\end{eqnarray}
where the $\tau^{x,z}$ are $4 \times 4$ matrices which
are block diagonal, the blocks consisting of $2\times 2$ Pauli
matrices. The velocities $v_{F,\Delta}$ are given by the conical
structure of $E_k$ near the $Q_{1-4}$: we have $v_F =
\left|\nabla_k \varepsilon_k |_{k=Q_a} \right|$ and $v_{\Delta} =
|J_1 \Delta_{x^2-y^2} \sqrt{2} \sin (Q)|$. In this limit, the energy of the
$\Psi_1$ fermionic excitations is
$E_k = ( v_F^2 (k_x+k_y)^2 /2 + v_\Delta^2 (k_x - k_y)^2 /2)^{1/2}$ (and similarly for $\Psi_2$), which is the spectrum of massless
Dirac fermions.

\subsection{Magnetic ordering}
\label{magd}

We now focus attention on the QPT involving loss of magnetic ordering within
the $d$-wave superconductor at $x=x_s$ in Fig.~\ref{fig:figglobal}. As in Section~\ref{sec:insulator},
we have to now consider the fluctuations of the SDW order parameter. We discussed two routes
to such a magnetic ordering transition in Section~\ref{sec:insulator}: one involving the vector SDW order
parameter in Section~\ref{sec:dimer}, and the other involving the spinor $z_\alpha$ in Section~\ref{sec:frust}.
In principle, both routes also have to be considered in the $d$-wave superconductor. The choice between
the two routes involves subtle questions on the nature of fractionalized excitations at intermediate scales
which we will not explore further here. These questions were thoroughly addressed in
Ref.~\cite{edoped} in the context of simple toy models: it was found that either route could
apply, and the choice depended sensitively on microscopic details. In particular, it was found that among
the fates of the non-magnetic superconductor was that it acquired VBS or Ising-nematic ordering,
as was found in the models explored in Section~\ref{sec:frust}. This is part of the motivation
for the expectation of such ordering in the regime $x_s < x < x_m$, as indicated in Fig.~\ref{fig:figglobal}.

In the interests of brevity and simplicity, we will limit our discussion of the SDW ordering transition
here to the vector formulation analogous to that in Section~\ref{sec:dimer}. We have full
square lattice symmetry, and so allow for incommensurate SDW ordering similar to the
$(q,q)$,$(q,-q)$ and $(\pi,q)$,$(q,\pi)$ states of Section~\ref{sec:frust}. Because there are two
distinct but degenerate ordering wavevectors, the complex order parameter $\Phi_a$ in Eq.~(\ref{sp2})
is now replaced by two complex order parameters $\Phi_{xa}$ and $\Phi_{ya}$ for orderings
along $(\pi, q)$ and $(q, \pi)$ (the orderings along $(q, \pm q)$ can be treated similarly
and we will not describe it explicitly). These order parameters are related to the spin operator by
\begin{equation}
S_a (\vec{r}) = \Phi_{xa} e^{i \vec{K}_x \cdot \vec{r}} + \Phi_{ya} e^{i \vec{K}_y \cdot \vec{r}} + \mbox{c.c.}
\label{phixy}
\end{equation}
where $\vec{K}_x = (q, \pi)$ and $\vec{K}_y = (\pi, q)$. As discussed below Eq.~(\ref{sp2}),
depending upon the structure of the complex numbers $\Phi_{xa}$, $\Phi_{ya}$, the SDW ordering
can be either collinear ({\em i.e.\/}, stripe-like) or spiral.
Also, as in Eqs.~(\ref{isingnematic1}) and (\ref{isingnematic2}), we can use these SDW order parameters
to also define a subsidiary Ising-nematic order parameter
\begin{equation}
\mathcal{I} =   |\Phi_{xa}|^2 - |\Phi_{ya}|^2 \label{isingnematic3}
\end{equation}
to measure the breaking of $x \leftrightarrow y$ symmetry.

Symmetry considerations will play an important role in our analysis of the
$\Phi_{x,ya}$ order parameters and their coupling to the Dirac fermions. In Table~\ref{table}
we therefore present a table of transformations under important operations of the square lattice
space group: these are easily deduced from the representations in Eq.~(\ref{g5a}) and
(\ref{phixy}).
\begin{table}[t]
\begin{center}
\begin{tabular}{c|ccccc}
& ~~~~~~~~$T_x$~~~~~~~~ &~~~~~~~~$T_y$~~~~~~~~  & ~~~~~~~~$R$~~~~~~~~ & ~~~~~~~~
$I$~~~~~~~~ & ~~~~~~~~$\mathcal{T}$~~~~~~~~\\
\hline
$\Phi_{xa}$ & $e^{iq} \Phi_{xa}$ & $- \Phi_{xa}$ & $\Phi_{ya}$ & $\Phi_{xa}^\ast$ & $- \Phi_{xa}$ \\
$\Phi_{ya}$ & $-  \Phi_{ya}$ & $e^{iq}\Phi_{ya}$ & $\Phi_{xa}^\ast$ & $\Phi_{ya}^\ast$ & $- \Phi_{ya}$ \\
$\Psi_{1 \alpha}$ & $e^{iQ} \Psi_{1\alpha}$ & $e^{iQ} \Psi_{1\alpha}$ &$ i \tau^z \Psi_{2\alpha}$ & $\Psi_{2\alpha}$ & $- \tau^y \Psi_{1\alpha}$\\
$\Psi_{2 \alpha}$ & $e^{-iQ} \Psi_{2\alpha}$ & $e^{iQ} \Psi_{2\alpha}$ &$ - i \varepsilon_{\alpha\beta} \left[ \Psi_{1\beta}^\dagger \tau^x \right]^T $
& $\Psi_{1\alpha}$ & $- \tau^y \Psi_{2\alpha}$\\
\end{tabular}
\end{center}
\vspace{0.1in}
\caption{
Transformations of the fields under operations which generate the symmetry group: $T_{x,y} = $ translation by a lattice
spacing in the $x,y$ directions, $R = $ rotation about a lattice site by 90$^\circ$, $I = $ reflection about the $y$ axis on a lattice site,
and $\mathcal{T} = $ time reversal. The theory is also invariant under spin rotations, with $i$ a vector index and $\alpha,\beta$ spinor indices.
We define $\mathcal{T}$ as an invariance of the imaginary time path integral, in which $\Phi_{1,2i}^\ast$ transform as the complex
conjugates of $\Phi_{1,2i}$, while $\Psi_{1,2\alpha}^\dagger$ are viewed as independent complex Grassman fields which transform
as $\Psi_{1,2\alpha}^\dagger \rightarrow \Psi_{1,2\alpha}^\dagger \tau^y$.
}
\label{table}
\end{table}

The effective action for the SDW order parameters has a direct generalization from (\ref{sp2}):
it can be obtained by requiring invariance under the transformations in Table~\ref{table}, and has many more allowed
quartic nonlinearities \cite{demler}:
\begin{eqnarray}
\mathcal{S}_{\Phi} &=& \int d^2 r d \tau \Biggl[ \left|
\partial_{\tau} \Phi_{x a} \right|^2 + c_x^2 \left|
\partial_{x} \Phi_{x a} \right|^2 + c_y^2 \left|
\partial_{y} \Phi_{x a} \right|^2 \nonumber \\ &~&~~~~~~~~+
 \left|
\partial_{\tau} \Phi_{y a} \right|^2 + c_x^2 \left|
\partial_{y} \Phi_{y a} \right|^2 + c_y^2 \left|
\partial_{x} \Phi_{y a} \right|^2 +
s \left( \left|\Phi_{x a}\right|^2 + \left|\Phi_{x a}\right|^2 \right) \nonumber \\
&~&~~~~~~~~+ \frac{u_1}{2}\left[  \left(
\left|\Phi_{x a} \right|^2 \right)^2 + \left(
\left|\Phi_{y a} \right|^2 \right)^2 \right]
+ \frac{u_2}{2}\left[  \left|
\left(\Phi_{x a} \right)^2 \right|^2  + \left|
\left(\Phi_{y a} \right)^2 \right|^2 \right] \nonumber \\ &~&~~~~~~~~+
  w_1 \left|\Phi_{x a}\right|^2 \left|\Phi_{y a}\right|^2
 +w_2 \left| \Phi_{xa} \Phi_{ya} \right|^2  +w_3 \left| \Phi_{xa} \Phi_{ya}^\ast \right|^2
 \Biggr]. \label{sp22}
\end{eqnarray}
Remarkably, a fairly complete 5-loop renormalization group analysis of this model has
been carried out by De Prato {\em et al.} \cite{vicari1}, and reliable information
on its critical properties is now available.

(We note parenthetically that Eq.~(\ref{sp22}) concerns the theory of the transition at $x_s$ from an SDW ordered state
to a $d$-wave superconductor with the full symmetry of the square lattice. However, as we have discussed
in Section~\ref{sec:intro} and in the beginning of Section~\ref{sec:dwave}, there could be Ising nematic
order in the regime $x_s < x < x_m$. In this case one of $\Phi_{xa}$ or $\Phi_{ya}$ orderings would
be preferred, and we need only consider the critical fluctuations of this preferred component. The
resulting action for this preferred component would then be identical to Eq.~(\ref{sp2}), with critical properties
as in Ref.~\cite{vicari0}.)

\begin{figure}[t]
\begin{center}
 \includegraphics[width=2in]{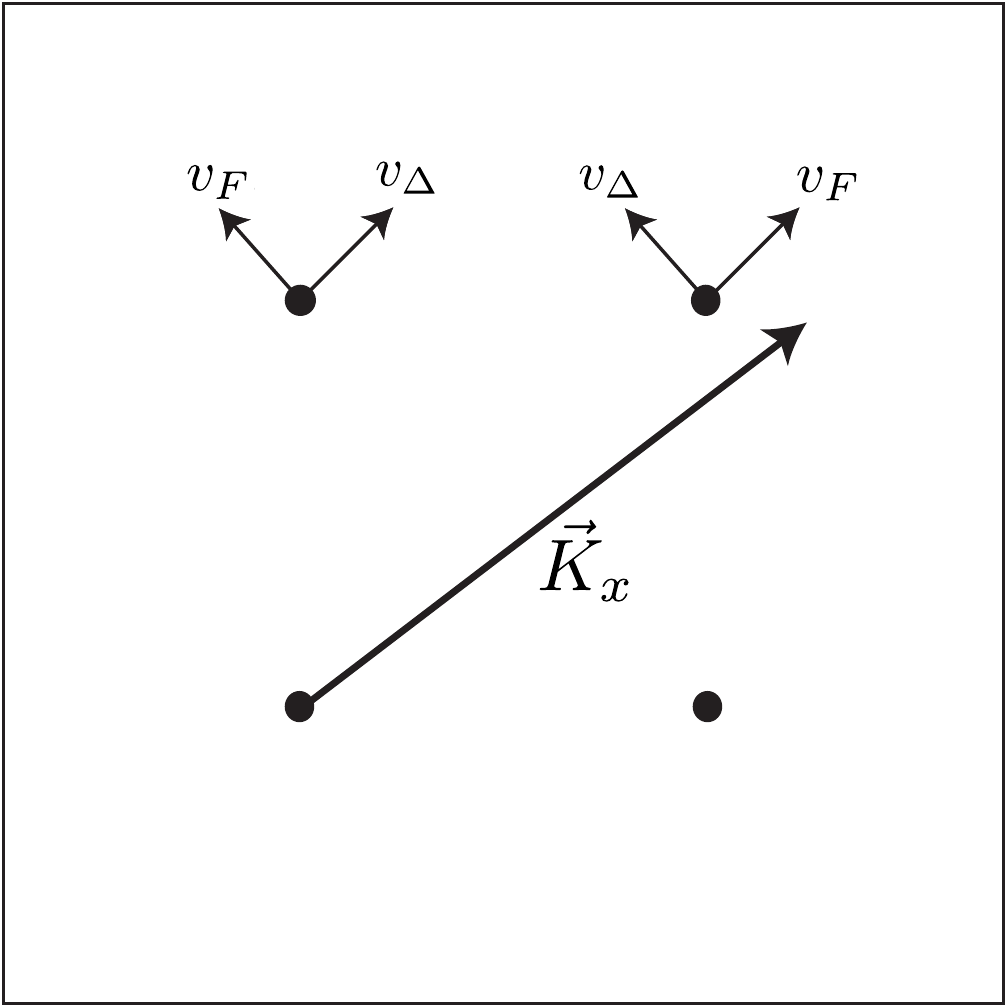}
 \caption{The filled circles indicate the positions of the gapless Dirac fermions in the
 square lattice Brillouin zone: these are at wavevectors $\vec{Q}_{1,2,3,4}$. An SDW fluctuation
 scatters a fermion at one of the nodes by wavevector $\vec{K}_x$ to a generic point in the Brillouin zone.
 The final state of the fermion has a high energy, and so such processes are suppressed.}
\label{fig:yukawa}
\end{center}
\end{figure}

Now we turn to the crucial issue of the coupling between the $\Phi_{x,ya}$ order parameter
degrees of freedom in $\mathcal{S}_\Phi$ and the massless Dirac fermions $\Psi_{1,2}$
in Eq.~(\ref{dsid1}). Again a great deal follows purely from symmetry considerations.
The simplest possible terms are cubic `Yukawa' interaction
terms like $\Phi_{xa} \Psi_{1}^\dagger \Psi_2$ etc.
However, these are generically forbidden by translational invariance, or equivalently, momentum
conservation. In particular, the transformation of the $\Phi_{x,ya}$ under translation by
one lattice spacing follows from (\ref{phixy}), while those of the $\Psi_{1,2}$ follow
from (\ref{g5a}). Unless the SDW ordering wavevectors $\vec{K}_{x,y}$ and
the positions of the Dirac nodes $\vec{Q}_{1,2,3,4}$ satisfy certain
commensurability conditions, the Yukawa coupling will not be invariant under this
translation operation. This is illustrated schematically in Fig.~\ref{fig:yukawa}.
The observed values of the wavevectors are not commensurate, and so we can safely neglect the Yukawa term.

The absence of the Yukawa coupling suggests that the fixed point theory describing the
QPT at $x=x_s$ in the superconductor may be $\mathcal{S}_\Phi$ in Eq.~(\ref{sp22})
alone, {\em i.e.\/}, the transition is in the same universality class as the insulator.
However, to ensure this, we have to examine the influence of higher terms coupling the degrees
of freedom of
$\mathcal{S}_\phi$ and $\mathcal{S}_\Psi$. The simplest couplings not prohibited by translational
invariance are associated with operators which are close to net zero momentum in both sectors.
These are further constrained by the other square lattice space group operations
in Table~\ref{table}; requiring invariance under them shows that
the simplest allowed terms are \cite{vicari2}
\begin{eqnarray}
&& \mathcal{S}_1 = \vartheta_1 \int d \tau d^2 r \left( |\Phi_{xa} |^2 + |\Phi_{ya}|^2 \right)
\left( \Psi^\dagger_{1} \tau^z \Psi_{1} +  \Psi^\dagger_{2} \tau^z \Psi_{2} \right) \nonumber \\
&& \mathcal{S}_2 = \vartheta_2 \int d \tau d^2 r \left( |\Phi_{xa} |^2 - |\Phi_{ya}|^2 \right)
\left( \Psi^\dagger_{1} \tau^x \Psi_{1} +  \Psi^\dagger_{2} \tau^x \Psi_{2} \right) \, . \label{evicari}
\end{eqnarray}
The first term is a fairly obvious `density-density' coupling between the energies of the two systems.
The second is more interesting: it involves
the Ising nematic order $\mathcal{I}$, as measured in the order parameter sector by (\ref{isingnematic3}), and in the fermion sector
by the bilinear shown above.

Now we can ask if the fixed point described by the decoupled theory $\mathcal{S}_\Phi + \mathcal{S}_\Psi$
is stable under the perturbations in $\mathcal{S}_1$ and $\mathcal{S}_2$. This involves a computation
of the scaling dimensions of the couplings $\vartheta_{1,2}$ at the decoupled theory fixed point.
These scaling dimensions were computed to 5-loop order in Ref.~\cite{vicari2}, and it was found that
$\mbox{dim}[\vartheta_1] \approx -1.0$, and $\mbox{dim}[\vartheta_2] \approx -0.1$. Thus both
couplings are irrelevant, and we can indeed finally conclude that the SDW onset transition is described
by the same theory as in the insulator. However, the scaling dimension of the $\vartheta_2$ coupling
is quite small, indicating that it will lead to appreciable effects. Thus we have demonstrated quite
generally that it is the Ising-nematic order $\mathcal{I}$ which is most efficient in coupling the SDW order
parameter fluctuations to the Dirac fermions. Note that $\mathcal{I}$ was not chosen
by hand, but was selected by the theory
among all other possible composite orders of the SDW field $\Phi_{x,ya}$.
The near zero scaling dimension of $\vartheta_2$ implies that it will induce a linewidth
$\sim T$ in the Dirac fermion spectrum. Moreover, because this broadening is mediated by $\mathcal{I}$, the broadening
will be strongly anisotropic in space \cite{kim}.

\subsection{Ising transitions}
\label{sec:ising}

Now we turn our attention to the vicinity of the point $x_m$ in Fig.~\ref{fig:figglobal}.
Although $x_m$ was defined in terms of the SDW transition in the metal at high magnetic
fields, we have also argued in Section~\ref{sec:intro} and in the beginning of Section~\ref{sec:dwave}
that there can also be transitions associated with VBS or Ising nematic order near $x_m$ but
within the superconducting phase at zero field. Strong evidence for a nematic order transition
near $x_m$ has emerged in recent experiments \cite{louisnematic,ando02,hinkov08a}.

This section will therefore consider the theory of Ising-nematic ordering within a $d$-wave
superconductor. Unlike the situation in Section~\ref{magd}, we will find here that the order parameter
and the Dirac fermions are strongly coupled, and the universality class of the transition is completely
changed by
the presence of the Dirac fermions. In Section~\ref{magd} we found that although the fermions
were moderately strongly coupled to the critical theory, they were ultimately reduced to
spectators to the asymptotic critical behavior.

Before considering the Ising nematic transition, we will take a short detour in Section~\ref{sec:did}
and describe another Ising transition associated with the breaking of time-reversal symmetry
in a $d$-wave superconductor. This leads to a model which has a somewhat simpler structure,
and for which conventional renormalization group techniques work easily. We will return to Ising-nematic
ordering in Section~\ref{sec:isingnematic}.

\subsubsection{Time-reversal symmetry breaking}
\label{sec:did}

We will consider a simple model in which the pairing symmetry of the
superconductor changes from $d_{x^2-y^2}$ to $d_{x^2-y^2} \pm i d_{xy}$.
The choice of the phase between the two pairing components
leads to a breaking of time-reversal symmetry. Studies of this transition
were originally motivated by the cuprate phenomenology, but we will not
explore this experimental connection here because the evidence has
remained sparse.

The mean field theory of this transition can be explored entirely within
the context of BCS theory, as we will review below. However, fluctuations
about the BCS theory are strong, and lead to non-trivial critical behavior
involving both the collective order parameter and the Bogoliubov fermions:
this is probably the earliest
known example \cite{vojta1,vojta2} of the failure of BCS theory in two (or higher)
dimensions in a superconducting ground state. At $T>0$, this
failure broadens into the ``quantum critical'' region.

We extend $H_{tJ}$ in Eq.~(\ref{g2}) so
that BCS mean-field theory permits a region with $d_{xy}$
superconductivity. It turns out that the frustrating interactions
as in Eq.~(\ref{hamil}) are precisely those needed. With a $J_2$ interaction,
Eq.~(\ref{g2}) is modified to:
\begin{equation}
\widetilde{H}_{tJ} = \sum_{k} \varepsilon_k c_{k \sigma}^{\dagger}
c_{k \sigma} + J_1 \sum_{\langle ij \rangle} \vec{S}_i \cdot {\bf
S}_{j} + J_2 \sum_{{\rm nnn}~ij} \vec{S}_i \cdot \vec{S}_j.
\label{g8}
\end{equation}
We will follow the evolution of the ground state of
$\widetilde{H}_{tJ}$ as a function of $J_2 / J_1$.

\begin{figure}
\centerline{\includegraphics[width=2in]{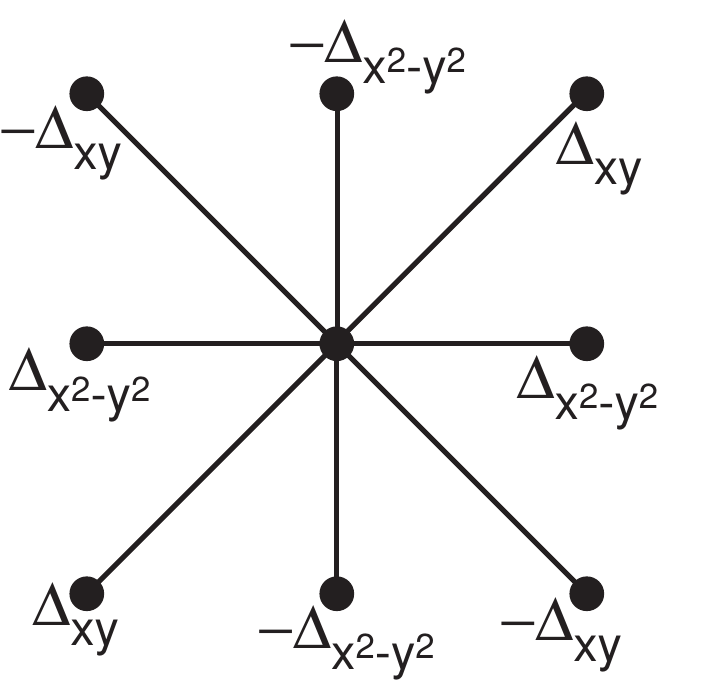}}
\caption{Values of the pairing amplitudes, $-\langle c_{i
\uparrow} c_{j \downarrow} -  c_{i \downarrow} c_{j \uparrow}
\rangle$ with $i$ the central site, and $j$ is one of its 8 nearest
neighbors.} \label{fig11}
\end{figure}

The
mean-field Hamiltonian is now modified from Eq.~(\ref{g3}) to
\begin{eqnarray}
\widetilde{H}_{BCS} = \sum_{k} \varepsilon_k c_{k
\sigma}^{\dagger} c_{k \sigma} &-& \frac{J_1}{2} \sum_{j,\mu}
\Delta_{\mu} (c_{j\uparrow}^{\dagger}
c_{j+\hat{\mu},\downarrow}^{\dagger} - c_{j\downarrow}^{\dagger}
c_{j+\hat{\mu},\uparrow}^{\dagger}) + \mbox{h.c.} \nonumber \\ &-&
\frac{J_2}{2} {\sum_{j,\nu}}^{\prime} \Delta_{\nu}
(c_{j\uparrow}^{\dagger} c_{j+\hat{\nu},\downarrow}^{\dagger} -
c_{j\downarrow}^{\dagger} c_{j+\hat{\nu},\uparrow}^{\dagger}) +
\mbox{h.c.} \, , \label{g9}
\end{eqnarray}
where the second summation over $\nu$ is along the diagonal
neighbors $\hat{x}+\hat{y}$ and $-\hat{x}+\hat{y}$. To obtain
$d_{xy}$ pairing along the diagonals, we choose $\Delta_{x+y} = -
\Delta_{-x+y} = \Delta_{xy}$. We summarize our choices for the
spatial structure of the pairing amplitudes (which determine the
Cooper pair wavefunction) in Fig.~\ref{fig11}.
The values of $\Delta_{x^2-y^2}$ and $\Delta_{xy}$ are to be
determined by minimizing the ground state energy (generalizing
Eq.~(\ref{g4}))
\begin{equation}
E_{BCS} = J_1 |\Delta_{x^2-y^2}|^2 +J_2 |\Delta_{xy}|^2 - \int
\frac{d^2 k}{4 \pi^2} \left[ E_k - \varepsilon_k \right] \, ,
\label{g10}
\end{equation}
where the quasiparticle dispersion is now (generalizing
Eq.~(\ref{g5}))
\begin{equation}
E_k = \left[ \varepsilon_k^2 + \left|J_1 \Delta_{x^2-y^2}(\cos k_x
- \cos k_y) + 2 J_2 \Delta_{xy} \sin k_x \sin k_y \right|^2
\right]^{1/2}. \label{g11}
\end{equation}
Notice that the energy depends upon the relative phase of
$\Delta_{x^2-y^2}$ and $\Delta_{xy}$: this phase is therefore an
observable property of the ground state.

\begin{figure}[t]
\centerline{\includegraphics[width=4in]{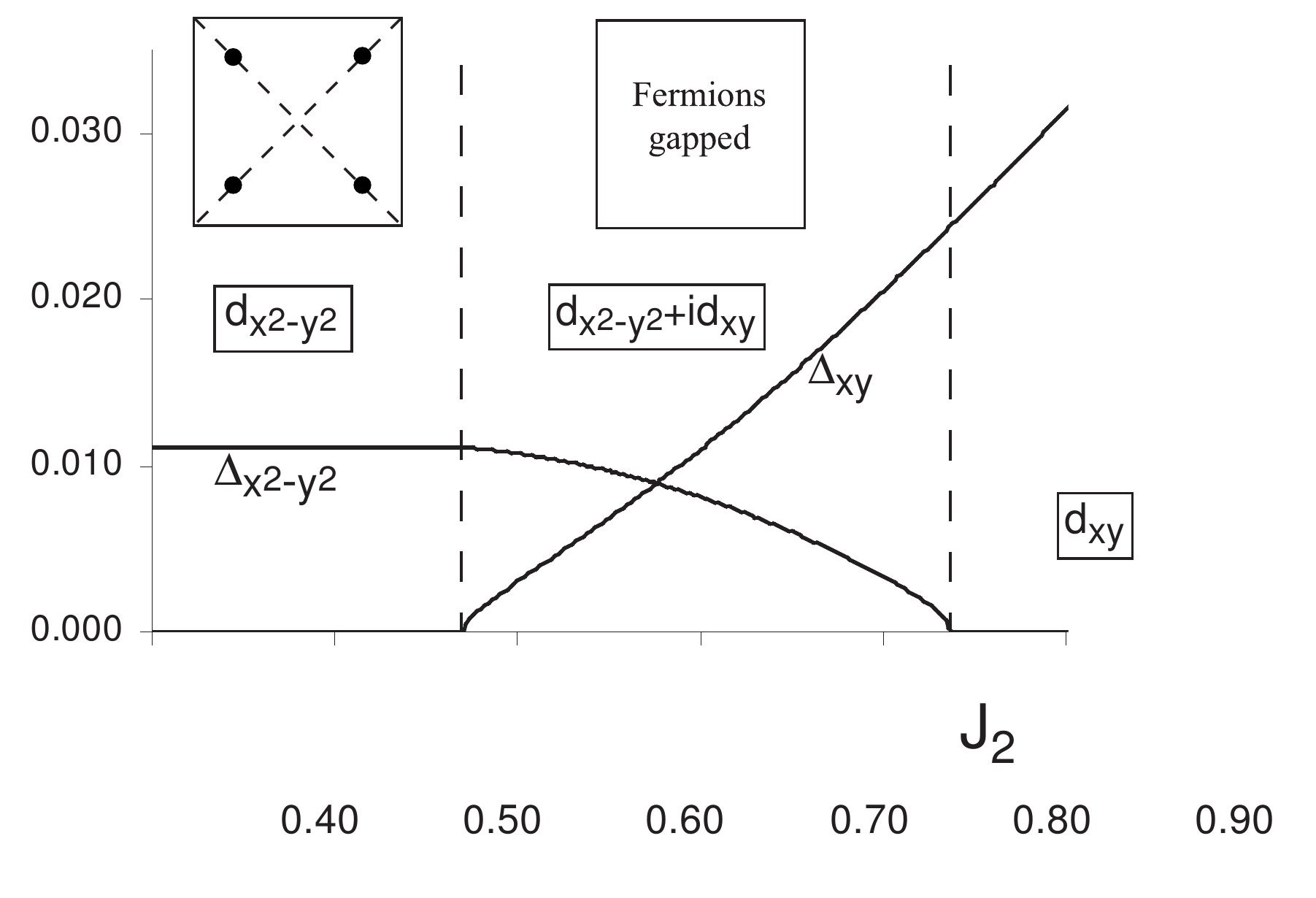}} \caption{BCS
solution of the phenomenological Hamiltonian $\widetilde{H}_{tJ}$
in Eq.~(\ref{g8}). Shown are the optimum values of the pairing
amplitudes $|\Delta_{x^2-y^2}|$ and $|\Delta_{xy}|$ as a function
of $J_2$ for $t_1 =1$, $t_2=-0.25$, $\mu=-1.25$, and $J_1$ fixed
at $J_1=0.4$. The relative phase of the pairing amplitudes was
always found to obey Eq.~(\ref{g12}). The dashed lines denote
locations of phase transitions between $d_{x^2-y^2}$,
$d_{x^2-y^2}+id_{xy}$, and $d_{xy}$ superconductors. The pairing
amplitudes vanishes linearly at the first transition
corresponding to the exponent $\beta_{BCS}=1$ in
Eq.~(\ref{g14}). The Brillouin zone location of the gapless Dirac points
in the $d_{x^2-y^2}$ superconductor is indicated by filled circles.
For the dispersion $\varepsilon_k$ appropriate to
the cuprates, the $d_{xy}$ superconductor is fully gapped, and so the
second transition is ordinary Ising. } \label{fig12}
\end{figure}

It is a simple matter to numerically carry out the minimization of
Eq.~(\ref{g11}), and the results for a typical choice of parameters
are shown in Fig.~\ref{fig12} as a function $J_2/J_1$.
One of the two amplitudes $\Delta_{x^2-y^2}$ or $\Delta_{xy}$ is
always non-zero and so the ground state is always superconducting.
The transition from pure $d_{x^2-y^2}$ superconductivity to pure
$d_{xy}$ superconductivity occurs via an intermediate phase in
which {\em both} order parameters are non-zero. Furthermore, in
this regime, their relative phase is found to be pinned to $\pm
\pi/2$, {\em i.e.}
\begin{equation}
\arg (\Delta_{xy}) = \arg (\Delta_{x^2-y^2}) \pm \pi/2 \label{g12} \, .
\end{equation}
The reason for this pinning can be intuitively seen from
Eq.~(\ref{g11}): only for these values of the relative phase does the
equation $E_k = 0$ never have a solution. In other words, the
gapless nodal quasiparticles of the $d_{x^2-y^2}$ superconductor
acquire a finite
energy gap when a secondary pairing with relative phase $\pm
\pi/2$ develops. By a level repulsion picture, we can expect that
gapping out the low-energy excitations should help lower the
energy of the ground state. The intermediate phase obeying
Eq.~(\ref{g12}) is called a $d_{x^2-y^2} + i d_{xy}$ superconductor.

The choice of the sign in Eq.~(\ref{g12}) leads to an overall two-fold
degeneracy in the choice of the wavefunction for the $d_{x^2-y^2}
+ i d_{xy}$ superconductor. This choice is related to the breaking
of time-reversal symmetry, and implies that the $d_{x^2-y^2} + i
d_{xy}$ phase is characterized by the non-zero expectation value
of a $Z_2$ Ising order parameter; the expectation value of this
order vanishes in the two phases (the $d_{x^2-y^2}$ and $d_{xy}$
superconductors) on either side of the $d_{x^2-y^2}+id_{xy}$
superconductor. As is conventional, we will represent the Ising
order by a real scalar field $\phi$. Fluctuations of $\phi$ become
critical near both of the phase boundaries in Fig.~\ref{fig12}. As
we will explain below, the critical theory of the $d_{x^2-y^2}$ to
$d_{x^2-y^2} + i d_{xy}$ transition is {\em not} the usual $\phi^4$ field
theory which describes the ordinary Ising transition in three
spacetime dimensions. (For the dispersion $\varepsilon_k$ appropriate to
the cuprates, the $d_{xy}$ superconductor is fully gapped, and so the
$d_{x^2-y^2} + i d_{xy}$ to $d_{xy}$ transition in Fig.~\ref{fig12} will be
ordinary Ising.)

Near the phase boundary from $d_{x^2-y^2}$ to $d_{x^2-y^2} +
id_{xy}$ superconductivity it is clear that we can identify
\begin{equation}
\phi = i \Delta_{xy}, \label{o1}
\end{equation}
(in the gauge where $\Delta_{x^2-y^2}$ is real). We can now expand
$E_{BCS}$ in Eq.~(\ref{g10}) for small $\phi$ (with $\Delta_{x^2-y^2}$
finite) and find a series with the structure~\cite{laugh,wolf}
\begin{equation}
E_{BCS} = E_0 + s \phi^2 + v |\phi|^3 + \ldots, \label{g13}
\end{equation}
where $s$, $v$ are coefficients and the ellipses represent regular
higher order terms in even powers of $\phi$; $s$ can have either
sign, whereas $v$ is always positive. Notice the non-analytic
$|\phi|^3$ term that appears in the BCS theory
-- this arises from an infrared singularity in the integral in
Eq.~(\ref{g10}) over $E_k$ at the four nodal points of the
$d_{x^2-y^2}$ superconductor, and is a preliminary indication that
the transition differs from that in the ordinary Ising model, and that the Dirac
fermions play a central role. We can optimize
$\phi$ by minimizing $E_{BCS}$ in Eq.~(\ref{g13}) -- this shows that
$\langle \phi \rangle =0$ for $s>0$, and $\langle \phi \rangle
\neq 0$ for $s<0$. So $s \sim (J_2/J_1)_c - J_2/J_1$ where
$(J_2/J_1)_c$ is the first critical value in Fig.~\ref{fig12}. Near
this critical point, we find
\begin{equation}
\langle \phi \rangle \sim (s_c - s)^{\beta}, \label{g14}
\end{equation}
where we have allowed for the fact that fluctuation corrections
will shift the critical point from $s=0$ to $s=s_c$. The present
BCS theory yields the exponent $\beta_{BCS} = 1$; this differs
from the usual mean-field exponent $\beta_{MF} = 1/2$, and this is
of course due to the non-analytic $|\phi|^3$ term in Eq.~(\ref{g13}).

We have laid much of the ground work for the required field theory
of the onset of $d_{xy}$ order
in Section~\ref{magd}. In addition to the order parameter $\phi$,
the field theory should also involve the low-energy nodal fermions
of the $d_{x^2-y^2}$ superconductor, as described by
$\mathcal{S}_{\Psi}$ in Eq.~(\ref{dsid1}). For the $\phi$ fluctuations,
we write down the usual terms permitted near a phase transition
with Ising symmetry, and similar to those in Eq.~(\ref{sp}):
\begin{equation}
\mathcal{S}_{\phi} = \int d^2 r d\tau \left[\frac{1}{2} \left(
(\partial_{\tau} \phi)^2 + c^2 (\partial_x \phi)^2 + c^2
(\partial_y \phi)^2 + s \phi^2 \right) + \frac{u}{24} \phi^4
\right]. \label{g15}
\end{equation}
Note that, unlike Eq.~(\ref{g13}), we do not have any non-analytic
$|\phi|^3$ terms in the action: this is because we have not
integrated out the low-energy Dirac fermions, and the terms in
Eq.~(\ref{g15}) are viewed as arising from high-energy fermions away
from the nodal points. Finally, we need to couple the $\phi$ and
$\Psi_{1,2}$ excitations. Their coupling is already contained in
the last term in Eq.~(\ref{g9}): expressing this in terms of the
$\Psi_{1,2}$ fermions using Eq.~(\ref{g5a}) we obtain
\begin{equation}
\mathcal{S}_{\Psi \phi} = \vartheta_{xy} \int d^2 r d \tau \left[  \phi
\left( \Psi_1^{\dagger} \tau^y \Psi_1 - \Psi_2^{\dagger} \tau^y
\Psi_2 \right) \right], \label{g16}
\end{equation}
where $\vartheta_{xy}$ is a coupling constant.
This coupling also has been obtained by symmetry considerations, by
examining invariants under the transformations of Table~\ref{table}.
The partition function of
the full theory is now
\begin{equation}
\mathcal{Z} = \int \mathcal{D}\phi \mathcal{D} \Psi_1 \mathcal{D}
\Psi_2 \exp \left( - \mathcal{S}_{\Psi} - \mathcal{S}_{\phi} -
\mathcal{S}_{\Psi \phi} \right), \label{g17}
\end{equation}
where $\mathcal{S}_{\Psi}$ was in Eq.~(\ref{dsid1}). It can now be
checked that if we integrate out the $\Psi_{1,2}$ fermions for a
spacetime independent $\phi$, we do indeed obtain a $|\phi|^3$
term in the effective potential for $\phi$.

We begin our analysis of $\mathcal{Z}$ in Eq.~(\ref{g17}) by following the
procedure of Section~\ref{magd}. Assume that the transition
is described by a fixed point with $\vartheta_{xy}=0$: then as in
Section~\ref{magd}, the theory for the transition would be the
ordinary $\phi^4$ field theory $\mathcal{S}_{\phi}$, and the nodal
fermions would again be innocent spectators.
The scaling
dimension of $\phi$ at such a fixed point is $(1 + \eta_I )/2$
(where $\eta_I$ is the anomalous order parameter exponent at the
critical point of the ordinary three dimensional Ising model),
while that of $\Psi_{1,2}$ is 1. Consequently, the scaling
dimension of $\vartheta_{xy}$ is $(1-\eta_I)/2 > 0$. This positive scaling dimension
implies that $\vartheta_{xy}$ is relevant and the $\vartheta_{xy}=0$ fixed point
is unstable: the Dirac fermions are fully involved in the critical theory.

Determining the correct critical behavior now requires a full
renormalization group analysis of $\mathcal{Z}$. This has been
described in some detail in Ref.~\cite{vojta2}, and we will not
reproduce the details here. The main result we need for our
purposes is that couplings $\vartheta_{xy}$, $u$, $v_F /c$ and
$v_{\Delta}/c$ all reach {\em non-zero} fixed point values which
define a critical point in a new universality class. These fixed
point values, and the corresponding critical exponents, can be
determined in expansions in either $(3-d)$~\cite{vojta1,vojta2}
(where $d$ is the spatial dimensionality) or $1/N$~\cite{kvesch}
(where $N$ is the number of fermion species).
An important simplifying feature here is that the fixed point is actually
relativistically invariant.
Indeed the fixed
point has the structure of the so-called Higgs-Yukawa model which
has been studied extensively in the particle physics
literature~\cite{baruch} in a different physical context: quantum
Monte Carlo simulation of this model also exist~\cite{qmc}, and
provide probably the most accurate estimate of the exponents.

The non-trivial fixed point has strong implications for
the correlations of the Bogoliubov fermions. The
fermion correlation function $G_1 = \langle \Psi_1
\Psi_1^{\dagger} \rangle$ obeys
\begin{equation}
G_1 (k, \omega) = \frac{\omega + v_F k_x \tau^z + v_{\Delta}
\tau^x}{(v_F^2 k_x^2 + v_{\Delta}^2 k_y^2 -
\omega^2)^{(1-\eta_f)/2}} \label{g18}
\end{equation}
at low frequencies for $s \geq s_c$. Away from the critical point
in the $d_{x^2-y^2}$ superconductor with $s>s_c$, Eq~(\ref{g18})
holds with $\eta_f = 0$, and this is the BCS result, with sharp
quasi-particle poles in the Green's function. At the critical
point $s=s_c$ Eq.~(\ref{g18}) holds with the fixed point values for
the velocities (which satisfy $v_F = v_{\Delta} = c$) and with the
anomalous dimension $\eta_f \neq 0$
-- the $(3-d)$ expansion~\cite{vojta1} estimate is $\eta_f
\approx (3-d)/14$, and the $1/N$ expansion estimate~\cite{kvesch}
is $\eta_f \approx 1/(3 \pi^2 N)$, with $N=2$. This is clearly
non-BCS behavior, and the fermionic quasiparticle pole in the
spectral function has been replaced by a branch-cut representing
the continuum of critical excitations. The corrections to BCS
extend also to correlations of the Ising order $\phi$: its
expectation value vanishes as Eq.~(\ref{g14}) with the Monte Carlo
estimate $\beta \approx 0.877$~\cite{qmc}. The critical point
correlators of $\phi$ have the anomalous dimension $\eta
\approx 0.754$~\cite{qmc}, which is clearly different from the
very small value of the exponent $\eta_I$ at the unstable
$\vartheta_{xy}=0$ fixed point. The value of $\beta$ is related to $\eta$
by the usual scaling law $\beta = (1+\eta) \nu/2$, with $\nu
\approx 1.00$ the correlation length exponent (which also differs
from the exponent $\nu_I$ of the Ising model).

\subsubsection{Nematic ordering}
\label{sec:isingnematic}

We now turn, as promised, to the case of Ising-nematic ordering within the $d$-wave superconductor
at $x=x_m$.

The ingredients of such an ordering are actually already present in our simple review of
BCS theory in Section~\ref{sec:dirac}. In Eq.~(\ref{g3}), we introduce 2 variational pairing amplitudes
$\Delta_x$ and $\Delta_y$. Subsequently, we assumed that the minimization of the energy
led to a solution with $d_{x^2-y^2}$ pairing symmetry with $\Delta_x = - \Delta_y = \Delta_{x^2 - y^2}$.
However, it is possible that upon including the full details of the microscopic interactions we
are led to a minimum where the optimal solution also has a small amount of $s$-wave pairing.
Then $|\Delta_x| \neq |\Delta_y|$, and we would expect all physical properties to have distinct
dependencies on the $x$ and $y$ coordinates. So, as in Eqs.~(\ref{isingnematic1}),
(\ref{isingnematic2}) and (\ref{isingnematic3}), we can define the Ising-nematic
order parameter by
\begin{equation}
\mathcal{I} = |\Delta_x|^2 - |\Delta_y|^2.
\label{isingnematic4}
\end{equation}

\begin{figure}[t]
\centerline{\includegraphics[width=3.5in]{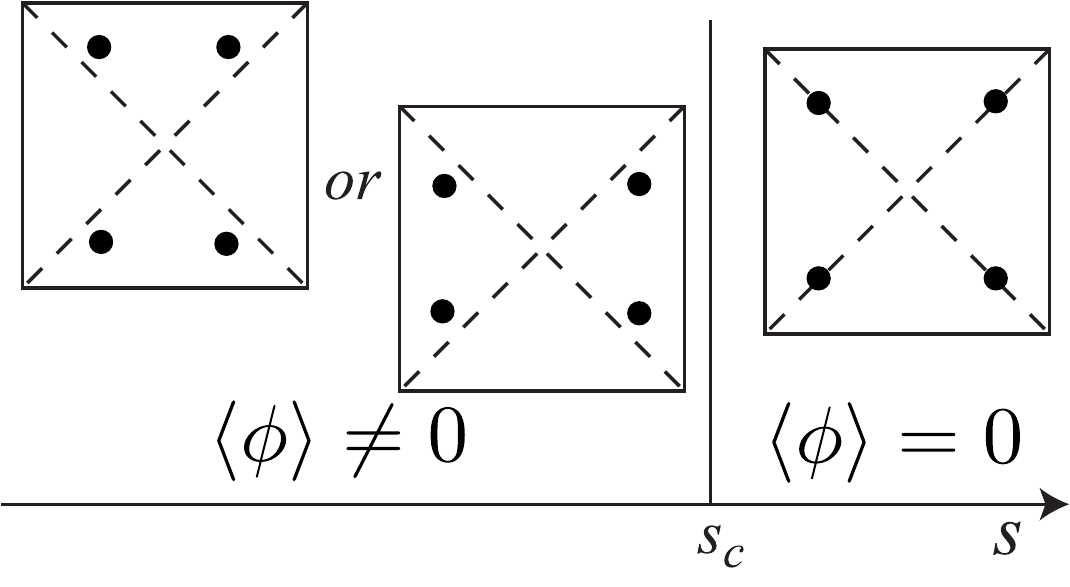}} \caption{
Phase diagram of Ising nematic ordering in a $d$-wave superconductor as a function
of the coupling $s$ in $\mathcal{S}_\phi$. The filled circles indicate the location
of the gapless fermionic excitations in the Brillouin zone. The two choices for $s<s_c$
are selected by the sign of $\langle \phi \rangle.$} \label{fig:dnematic}
\end{figure}

The derivation of the field theory for this transition follows closely our presentation in
Section~\ref{sec:did}. We allow for small Ising-nematic ordering by introducing a scalar field $\phi$
and writing
\begin{equation}
\Delta_x = \Delta_{x^2 - y^2} + \phi~~~;~~~\Delta_y = - \Delta_{x^2-y^2} + \phi ;
\end{equation}
note that $\mathcal{I} \propto \phi$. The evolution of the Dirac fermion spectrum under such
a change is indicated
in Fig.~\ref{fig:dnematic}.
We now develop an effective action for $\phi$ and the Dirac fermions $\Psi_{1,2}$.
The result is essentially identical to that in Section~\ref{sec:did}, apart from a change
in the structure of the Yukawa coupling. Thus we obtain a theory
$\mathcal{S}_\Psi + \mathcal{S}_\phi + \overline{\mathcal{S}}_{\Psi\phi}$,
defined by Eqs.~(\ref{dsid1}) and (\ref{g15}), and
where
Eq.~(\ref{g16}) is now replaced by
\begin{equation}
\overline{\mathcal{S}}_{\Psi \phi} = \vartheta_{I} \int d^2 r d \tau \left[  \phi
\left( \Psi_1^{\dagger} \tau^x \Psi_1 + \Psi_2^{\dagger} \tau^x
\Psi_2 \right) \right]. \label{h16}
\end{equation}
Not surprisingly, the fermion bilinear coupling to the nematic order parameter $\phi$
is identical to that in Eq.~(\ref{evicari}), as is expected from the transformations of Table~\ref{table}.

The seemingly innocuous change between Eqs.~(\ref{g16}) and (\ref{h16}) however
has strong consequences. This is partly linked to the fact with $\overline{\mathcal{S}}_{\Psi\phi}$
cannot be relativistically invariant even after all velocities are adjusted to be equal.
A weak-coupling renormalization group analysis in powers of the coupling $\vartheta_I$ was performed
in $(3-d)$ dimensions in Refs.~\cite{vojta1,vojta2}, and led to flows to strong coupling with no
accessible fixed point: thus no firm conclusions on the nature of the critical theory
were drawn.

This problem remained unsolved until the recent works of Refs.~\cite{kim,yejin}.
It is essential that the coupling $\vartheta_I$ not be used as a perturbative expansion parameter.
This is because it leads to strongly non-analytic changes in the structure of the $\phi$
propagator, which have to be included at all stages. In a model with $N$ fermion flavors,
the $1/N$ expansion does avoid any expansion in $\vartheta_I$. The renormalization group
analysis has to be carried out within the context of the $1/N$ expansion, and this
involves some rather technical analysis which is explained in Ref.~\cite{yejin}.
In the end, an asymptotically exact description of the vicinity of the critical point was obtained.
It was found that the velocity ratio $v_F / v_\Delta$ diverged logarithmically with energy scale,
leading to strongly anisotropic `arc-like' spectra for the Dirac fermions. Associated singularities
in the thermal conductivity have also been computed \cite{lars}.

\section{Metals}
\label{sec:metal}

We finally turn to the transition in the metal at $x_m$, which anchored our discussion
of the cuprate phase diagram in Section~\ref{sec:intro}. This controls the high field transition
line in Fig.~\ref{fig:figglobal} between the large Fermi surface and small Fermi pocket states.
We also argued that this transition was a key ingredient in a theory of the strange metal.

In addition to the order parameter ingredients we met in Section~\ref{sec:insulator},
we now have to also account for fermion excitations as in Section~\ref{sec:dwave}.
In Section~\ref{sec:dwave} the fermionic excitations had vanishing energy only
at isolated nodal points in the Brillouin zone: see Fig.~\ref{fig:yukawa}. In the present
section we are dealing with metals, which have fermionic excitations with vanishing energy
along an entire line in the Brillouin zone. Thus we can expect them to have an even
stronger effect on the critical theory. This will indeed be the case, and we will be led to problems
with a far more complex structure. Unlike the situation in insulators and $d$-wave superconductors,
many basic issues associated with ordering transitions in two dimensional metals have not been
fully resolved. The problem remains one of active research and is being addressed by
many different approaches.

As discussed in Sections~\ref{sec:insulator} and~\ref{magd}, we can describe magnetic ordering
by using either vector or spinor variables for the order parameter, and these lead to very different
phases and critical points. For metals,
the relationship between these two approaches, and their distinct physical properties
have been described recently in Ref.~\cite{su2acl}. The spinor route is more `exotic' and leads to
intermediate non-Fermi liquid critical phases between the small and large Fermi surface Fermi liquid phases.
These intermediate critical phases could well be important for the experiments and for Fig.~\ref{fig:figglobal}, but we will
not describe them here. We will limit our present discussion to the more conventional vector mode
description of the SDW ordering transition.

In  recent papers \cite{max1,max2} Metlitski and the author have argued
that the problem of symmetry breaking transitions in two-dimensional metals
is strongly coupled, and proposed field theories and scaling structures
for the vicinity of the critical point. Here we will be satisfied with a simple description of the effective
action and its mean field theory \cite{vojtarev}:
the reader is referred to the recent papers \cite{max1,max2} for
further analyses.

\begin{figure}[t]
\centering
 \includegraphics[width=4.6in]{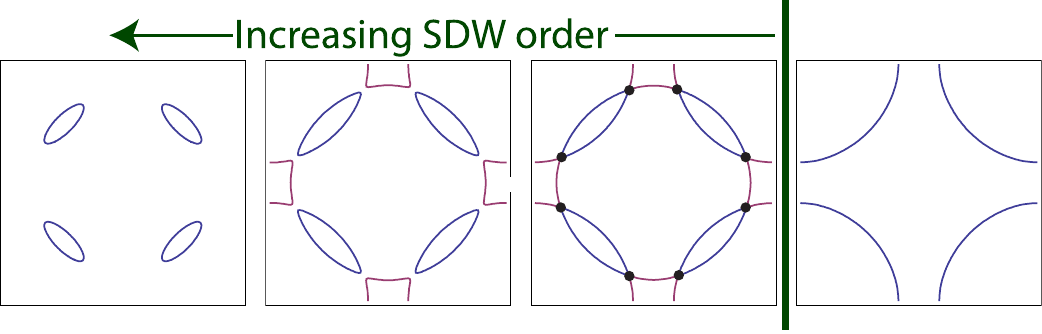}
 \caption{Evolution of the Fermi surface of the hole doped cuprates in a conventional SDW theory \cite{scs}
 as a function of the magnitude of the SDW order $|\varphi_a|$, obtained from Eq.~(\ref{sdwband}).
 The right panel
is the large Fermi surface state with no SDW order, with states contiguous to $\vec{k}=0$ occupied by electrons.
The ``hot spots'' are indicated by the filled circles in the second panel from the right.
The onset of SDW order induces the formation of electron and hole pockets (the hole pockets are the ones intersecting
the diagonals of the Brillouin zone). 
With further increase of $|\varphi_a|$, the electron pockets disappear and only
hole pockets remain (the converse happens in the last step for the electron-doped cuprates).}
\label{fig:sdw}
\end{figure}

As in Section~\ref{sec:dwave}, let us begin by a description of the non-critical fermionic
sector, before its coupling to the order parameter fluctuations.
We use the band structure describing the cuprates in the over-doped region, well away from the Mott insulator. Here the electrons
$c_{\vec{k} \alpha}$ are described by the kinetic energy in Eq.~(\ref{g2}), which we write in the
following action
\begin{equation}
\mathcal{S}_c = \int d\tau \sum_{\vec{k}} c_{k \alpha}^\dagger \left( \frac{\partial}{\partial \tau} + \varepsilon_k
\right) c_{k \alpha}. \label{sc}
\end{equation}
This band structure leads to the Fermi surface shown in the
right-most panel of Fig.~\ref{fig:sdw}, and also later in Fig.~\ref{fig:fnematic}.

\subsection{SDW ordering}
\label{sec:msdw}

As discussed in Section~\ref{sec:dwave}, we must now couple the fermions of Eq.~(\ref{sc}) to the bosonic
modes associated with the SDW ordering transition.
As noted above, we will use the more conventional vector mode
description of the SDW ordering transition using the order parameters in
Eq.~(\ref{phixy}). The analog of the coupling in Eq.~(\ref{g16})
now leads to the interaction term
\begin{equation}
\mathcal{S}_{c \Phi} = \int d \tau \sum_{\vec{k}, \vec{q}} \Phi_{xa} (\vec{q}) c_{\vec{k} + \vec{K}_x + \vec{q}, \alpha}^\dagger \sigma^a_{\alpha \beta} c_{k \beta} + \mbox{c.c.} + ~~x \rightarrow y
\, ,
\label{scp}
\end{equation}
where $\vec{q}$ is a small momentum associated with a long-wavelength SDW fluctuation,
while the sum over the momentum $\vec{k}$ extends over the entire Brillouin zone.
The complete theory for the SDW transition is now contained, in principle, in $\mathcal{S}_c +
\mathcal{S}_{c\Phi} + \mathcal{S}_\Phi$, with $\mathcal{S}_\Phi$ contained in Eq.~(\ref{sp22}).

Let us consider the mean-field predictions of this theory in the SDW ordered state.
For simplicity, we consider ordering at $\vec{K} = (\pi,\pi)$, in which case $\Phi_{xa}$
and $\Phi_{ya}$ both reduce to the N\'eel order $\varphi_a$ in Eq.~(\ref{sp}).
In the state with SDW order, we can take $\varphi_a = (0,0, \varphi)$ a constant.
Then $\mathcal{S}_c + \mathcal{S}_{c \Phi}$ is a bilinear in the fermions and can be diagonalized
to yield a fermion band structure (the analog of Eq.~(\ref{g5}))
\begin{equation}
E_{k} = \frac{\varepsilon_k + \varepsilon_{\vec{k} + \vec{K}}}{2} \pm
\left( \left(\frac{\varepsilon_k + \varepsilon_{\vec{k} + \vec{K}}}{2}\right)^2
+ \varphi^2 \right)^{1/2} \, .
\label{sdwband}
\end{equation}
Filling the lowest energy bands of this dispersion leads to the Fermi surface structure \cite{scs}
shown in Fig.~\ref{fig:sdw}.
The second panel from the right shows the Fermi surface obtained
by translating the original Fermi surface by $\vec{K}$, and the remaining panels show the consequences of mixing between the states
at momentum $\vec{k}$ and $\vec{k} + \vec{K}$.
Note that the Fermi surface has split apart into ``small'' electron and hole pockets, as discussed
in Section~\ref{sec:intro}.

Let us now attempt to move beyond this simple mean field theory. As written, the action  $\mathcal{S}_c +
\mathcal{S}_{c\Phi} + \mathcal{S}_\Phi$ is not conducive to a field-theoretic analysis: this is mainly because
the sum over $\vec{k}$ in Eq.~(\ref{scp}) extends over the entire Brillouin
zone, and there are low-energy
fermionic excitations along an entire line of $\vec{k}$ close the the Fermi surface. However, one simplifying
feature here is that most of these low-energy fermions do not couple efficiently to the SDW order parameter,
and their situation is similar to the fate of the Dirac fermions illustrated in
Fig.~\ref{fig:yukawa} -- upon scattering
with the wavevector $\vec{K}$, they end up at generic points in the Brillouin zone at which there are only
high-energy fermionic states. There are now 8 special ``hot spots'' on the Fermi surface which do connect
via the wavevector $\vec{K}$ to other spots directly on the Fermi surface: these are illustrated in Fig.~\ref{fig:sdw}.
These ``hot spots'' are thus similar to the Dirac hot spots we met in Section~\ref{sec:ising}
upon considering Ising transitions with
a zero-momentum order parameter in a $d$-wave superconductor. However the present situation is more
complex because we also have ``cold lines'' of zero energy fermionic excitations coming into the hot spots.

A successful theory of the fermionic hot spots was reviewed in Section~\ref{sec:ising}. A natural idea
is to apply the same approach to the present situation with fermionic hot spots and cold lines.
This leads to a problem of considerably complexity, which remains strongly coupled even within the context
of the $1/N$ expansion: see Refs.~\cite{max2,max3} for further details.

\subsection{Nematic ordering}
\label{sec:mnematic}

\begin{figure}[t]
\centerline{\includegraphics[width=3.5in]{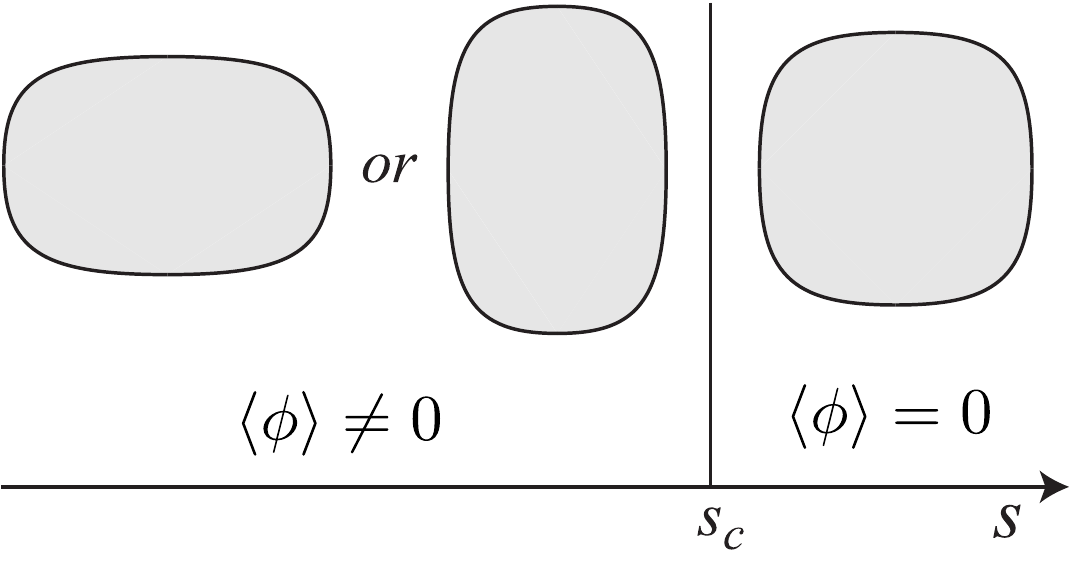}} \caption{
Phase diagram of Ising nematic ordering in a metal as a function
of the coupling $s$ in $\mathcal{S}_\phi$. The Fermi surface for $s>0$ is the same
as that in the right-most panel of Fig.~\ref{fig:sdw}, but with the $\vec{k}=0$ point shifted from
the center to the edge of the Brillouin zone. The interior regions are the occupied hole (or empty electron) states.
The choice between the two quadrapolar distortions
of the Fermi surface is determined by the sign of $\langle \phi \rangle$.} \label{fig:fnematic}
\end{figure}

For completeness, we also consider the case of the Ising-nematic ordering in the presence of the large
Fermi surface metal. Then we will have an Ising order parameter represented by the real scalar field $\phi$,
which is described as before by Eq.~(\ref{g15}). Its coupling to the electrons can be deduced by
symmetry considerations, and the most natural coupling (the analog of Eqs.~(\ref{h16}) and (\ref{scp})) is
\begin{equation}
\mathcal{S}_{c \phi} = \int d \tau \sum_{\vec{k}, \vec{q}} (\cos k_x - \cos k_y ) \phi (\vec{q}) c_{\vec{k}+\vec{q}/2, \alpha}^\dagger
c_{\vec{k}-\vec{q}/2, \alpha}.
\label{sci}
\end{equation}
The momentum dependent form factor is the simplest choice with changes sign under $x \leftrightarrow y$, as is required
by the symmetry properties of $\phi$. Again, the sum over $\vec{q}$ is over small momenta, while that over $\vec{k}$
extends over the entire Brillouin zone. The theory for the nematic ordering transition is now
described by $\mathcal{S}_c + \mathcal{S}_\phi +  \mathcal{S}_{c \phi}$. The evolution of the Fermi
surface as a function of the Ising coupling in $\mathcal{S}_\phi$ is shown in Fig.~\ref{fig:fnematic}.

Note that Eq.~(\ref{sci}) does not have any large momentum transfer associated with $\vec{K}$. Consequently, at any generic
point on the Fermi surface, there can be scattering to other nearby low-energy fermionic excitations by long wavelength
modes of $\phi$. In other words, the entire Fermi surface is ``hot''. Thus we are faced with a third case of a
``hot line'' of fermions coupled to the critical order parameter mode of the transition. This case has been
analyzed in Ref.~\cite{max1}, where it is proposed that the critical point is actually described by
an infinite number of 2+1 dimensional field theories, labeled by points on the Fermi surface. The reader is referred
to Ref.~\cite{max1} for further results on this complex problem -- a review of the main results appears
in Ref.~\cite{ssaegean}.

\subsubsection*{Acknowledgements}

I thank Eun Gook Moon for valuable comments on the manuscript and for a collaboration \cite{moon} which led
to Fig.~\ref{fig:figglobal}, R.~Fernandes, J.~Flouquet, G.~Knebel, and J.~Schmalian,
for providing the plots shown in Fig.~\ref{fig:knebel}, C.~Ruegg for the plot shown in Fig.~\ref{fig:ruegg},
and the participants of the schools for their interest, and for stimulating discussions.
This research was supported by the National Science Foundation under grant DMR-0757145, by the FQXi
foundation, and by a MURI grant from AFOSR.

\end{document}